\documentclass[a4paper, 11pt]{article}
\pdfoutput=1

\usepackage{jheppub}
\usepackage{subcaption}
\usepackage{float}
\usepackage{graphicx}
\usepackage{xcolor}
\usepackage{mathtools}

\usepackage{physics}
\usepackage{amsmath,amsthm,amssymb}
\usepackage{natbib}
\usepackage[font=small,labelfont=bf]{caption}
\usepackage{bm}

\usepackage{hyperref}% add hypertext 

\renewcommand{\v}[1]{ \ensuremath{ {\bm{#1}} }}

\usepackage{enumitem,amssymb}
\newlist{todolist}{itemize}{2}
\setlist[todolist]{label=$\square$}
\usepackage{pifont}

\newcommand{\DDelta}[2]{ \ensuremath{\delta^{(2)} (#1-#2)  }}

\title{CGC for Ultra-Peripheral Pb+Pb Collisions at the Large Hadron Collider: a more realistic calculation.}

\author[a]{Haowu Duan,}
\affiliation[a]{Department of Physics, North Carolina State University, Raleigh, NC 27695, USA}

\author[b]{Alexander Kovner,}
\affiliation[b]{Physics Department, University of Connecticut, 2152 Hillside Road, Storrs, CT 06269, USA}

\author[a,c]{and Vladimir V. Skokov}
\affiliation[c]{RIKEN-BNL Research Center, Brookhaven National Laboratory, Upton, NY 11973, USA}

\abstract{
We provide the first calculation of two-gluon production at mid-rapidity in ultra-peripheral collisions  in the Color Glass Condensate framework. To estimate systematic uncertainty associated with poor understanding of  the wave function of the nearly real photon, we consider two diametrically different models: the dilute quark-antiquark dipole approximation and a 
vector meson, in which color charge density is approximated by  McLerran-Venugopalan model.  In the experimentally relevant range, the target nucleus can be faithfully approximated by a highly saturated state. This simplification enables us to perform efficient numerical simulations and extract the two-gluon correlation functions and the associated azimuthal harmonics.      
}

\usepackage{hyperref}
\hypersetup{
    colorlinks,
    citecolor=blue,
    filecolor=blue,
    linkcolor=blue,
    urlcolor=blue
}

\date{\today}

\begin{document}
\maketitle
%\tableofcontents
\newpage

\section{Introduction} 
Deep inelastic scattering off a nuclear target at high energies  constitutes the best environment to  probe the small x tail of the hadron wave function and  to  confirm  experimentally the existence of gluon saturation. 
There is a trivial reason why one is interested in the smallest probe possible: to minimize the effect of final state interaction and make  experimental data interpretable from the perspective of the hadron wave function. At high energy, the coherence length of the virtual photon is significantly larger than the proton radius. Thus the larger the nuclear target the denser the gluon field which  the projectiles probes. While the Electron Ion Collider will be the best discovery machine for the saturation physics,  the ultra-peripheral collisions~\cite{Bertulani:2005ru,Baltz:2007kq} at RHIC and the LHC may also be a very useful probe of the small x hadron wave function as a nearly real photon likely has a smaller partonic content than the proton, and consequently the influence of final state effects should be smaller as well. 

In this paper we focus on correlations between produced particles in Ultraperipheral Collisions (UPC). Traditionally, hydrodynamics has been successful in dealing with correlations produced in heavy ion collisions. However, the discovery of such correlation in small systems brought to the fore the question of their origin. Following the high multiplicity pp measurement by CMS~\cite{CMS:2010ifv}, a set of measurements was done in the past decade, specifically, p-Pb at LHC~\cite{ALICE:2012eyl,ATLAS:2012cix,CMS:2012qk} and p-Au, d-Au, He-Au at RHIC~\cite{PHENIX:2016cfs,PHENIX:2013ktj,STAR:2014qsy}, see more details in review articles~\cite{Schenke:2021mxx,Nagle:2018nvi}. In addition similar correlations have been recently observed in UPC \cite{ATLAS:2021jhn}. Similar measurement was also performed in $e^+e^-$ collision \cite{Badea:2019vey}, and ep Deep-inelastic-scattering \cite{ZEUS:2019jya}, and no ridge correlation was found in this data. 

Our focus in this paper is on the possibility that particle correlations originate from the quantum correlation in the initial state hadronic wave function.
From theoretical perspective, the wave-function of a nearly real photon is poorly understood due to uncontrollably large non-perturbative effects. Thus theoretical results describing scattering and particle production in UPC are unavoidably model dependent. In this paper, in order to at least provide a sense of the systematic uncertainty, we consider two models for the photon wave functions. The first one is based on a naive perturbative process of photon splitting into a virtual quark-antiquark pair. The second assumes that the real photon can be represented as, in general non-perturbative, hadronic state, e.g. $\rho$-meson~\cite{Sakurai:1960ju}. A $\rho$-meson wave-function evolved to asymptotically high energy/small x would be similar to the wave function of a nucleus or a proton and thus can be approximated by the McLerran-Venugopalan (MV) model. For the energies available at the  LHC and especially at RHIC, this approximation can be expected to be wanting. Nonetheless, two diametrically opposite approximations may help to understand the influence and the importance of the associated systematics.   

Within the Color Glass Condensate approach, the scattering and central rapidity particle production in photon-nucleus collisions pose yet another challenge (see Ref.~\cite{Zhao:2022ayk} for hydrodynamic approach to UPC).  
At the moment, the two - particle production was only studied in the forward, photon going direction~\cite{Shi:2020djm}, which is not appropriate to describe the central rapidity production for two reasons. First, theoretically, forward production is a significantly different process than the production at central rapidity and as such it has qualitatively  different systematics as a function of the control parameters. Second,  the experimental data shows extremely strong rapidity dependence, see Ref.~\cite{ATLAS:2021jhn}. The lack of boost-invariance~\cite{Bozek:2015swa,Ke:2016jrd,Schenke:2016ksl,Shen:2020jwv,Wu:2021hkv} invalidates approximating of the mid-rapidity physics by the physics of the forward region. In particular only a small fraction of particles is produced in the forward region. Thus it is hard to imagine that particle correlations integrated over a sizable rapidity interval can be reasonably described in terms of these hadrons alone as in \cite{Shi:2020djm}. Thus addressing mid-rapidity production is absolutely crucial in order for the CGC based theoretical analysis to be taken seriously. 

Here we undertake such a calculation, albeit within the constraints mentioned above.
As in the rest of the CGC based calculations, we consider inclusive two-gluon production from the projectile wave function in the eikonal approximation for the emission vertex. This approximation is justified if the gluons have sufficiently different rapidities. On the other hand when the rapidity difference becomes too large, $\sim \frac{1}{\alpha_s}$ this approximation has to be revised as one would be required to account for the evolution between the gluon emissions~\cite{Jalilian-Marian:2004vhw}. However experimentally one is not dealing with such large rapidity differences, and so our simple approximation should be adequate. 

The calculation of particle production in the CGC framework necessitates averaging of products of Wilson lines over the color field ensemble of the target. 
This in general is a rather complex endeavor since to study correlations we are forced to go beyond the large $N_c$ limit which then requires us to calculate correlators of a large number of Wilson lines. In order to make the calculations tractable and not overly numerically demanding, in this paper we use the so-called {\it factorized  dipole approximation} (FDA) discussed in Ref.~\cite{Kovner:2017ssr,Kovner:2018vec} and successfully applied to study particle production and correlations in Refs.~\cite{Altinoluk:2018ogz,Kovner:2018fxj,Kovner:2018azs,Li:2021ntt,Altinoluk:2020psk}. This approximation is well suited for a dense nuclear target in and close to saturation regime. We will explain this in more detail in the following.

%The complexity of the used approach is significantly more severe than in the forward direction studies of Ref.~\cite{Shi:2020djm}, as the reader will be discover in the following section. Nonetheless, our work is not a final word on this topic given  the necessary approximations we listed  above. 
The paper is organized as follows. 
We start from summarizing the machinery used to calculate particle production in dilute-dense scattering in Sec.~\ref{sec2}. In Sec.~\ref{sec3}, we calculate a general expression for double gluon production cross section. In Sec.~\ref{sec4} and Sec.~\ref{sec5}, we discuss projectile ensemble average for the dipole and MV models respectively.  Sec.~\ref{Sec:Num} lays out the details of our numerical procedure. We conclude with the discussion of our results in  Sec.~\ref{sec7}.

%-------------------------
\section{Particle production in dilute-dense scattering}\label{sec2}
In this section we outline the most important ingredients of the formalism and define the observables. 
The formalism of our choice is the wave function approach of Refs.~\cite{Kovner:2001vi,Baier:2005dv}. Alternatively one can use diagrammatic methods of Ref.~\cite{Kovchegov:2012mbw}. While the final expressions are the same, the representation we use is convenient for organizing calculations.  

%Here we review the formalism discussed in Ref.~\cite{Baier:2005dv}. 

Consider a projectile at high energy before a collision with a hadronic/nuclear target. The projectile's wave function can be written in the general form
\begin{align}
    | \Psi_{\rm in} \rangle  = 
    \sum_{\{\v{x}, k^+, \alpha\}} 
    \psi (\{\v{x}, k^+, \alpha\})
   \prod_i|{\v{x}_i, k^+_i, \alpha_i}\rangle 
\end{align}
where $\v{x}_i$ is the $i$-th parton's  transverse position, $k_i^+$ longitudinal momentum and  $\alpha_i$ the color index. Subject to the physical process, the latter can belong to either (anti)fundamental 
or adjoint representation describing a (anti)quark or a gluon in the incoming wave function. %The sum runs over all variables.  

%The general framework of scattering can be extremely complicated. 
At high energy, the separation of relevant timescales allows for significant simplifications of the analysis of the scattering process. The propagation time through the target in high energy collisions can be arbitrary small compared to effective inter-parton interaction time. One thus can neglect the interactions between the partons and approximate the scattering process by independent propagation of partons.
 For a similar reason, 
at high energy 
%interaction can be considered in the eikonal approximation according to which 
the partons do not change their transverse positions during the scattering process. The high energy scattering in QCD is non-trivial due to the different eikonal  phases acquired by different components of the wave function. This difference in phases is responsible for decoherence of the components of the wave function and leads to particle production in the final state.  The wave function of the outgoing projectile right after scattering is
\begin{align}\label{out}
    | \Psi_{\rm out} \rangle  = S |\Psi_{\rm in}\rangle = \sum_{\{\v{x}, k^+, \alpha\}} 
    \psi (\{\v{x}, k^+, \alpha\})
    \prod_i U_{\alpha_i \beta_i}( \v{x}_i)    |\v{x}_i, k^+_i, \beta_i\rangle 
\end{align}
where  $U$ is the Wilson line defined in the representation of $SU(N_c)$ group appropriate to the color charge of a given parton: 
\begin{align}
    U_{\v{x}}  = {\cal P} e^{ i \int d x^- T^a A^+_a(\v{x},x_-) }.
\end{align}
Here $A^+$ is the $+$ component of the gauge field of the target.   

To calculate the expectation value of a gluon observable %by $ O(a,a^\dagger) $ 
in the final state, one has to allow for the propagation of the projectile state \eqref{out} to asymptotic time $t\rightarrow\infty$.
This evolution results in emission of additional gluons which dress the bare partons by their WW fields as well as in recombination of some of the outgoing soft gluons into the WW field of the outgoing fast partons. %This is nothing else but the high energy counterpart of the Faddeev-Kulish construction  of dressed states in theories with massless gauge bosons. 
It is straightforward to account for the evolution after scattering (or equivalently to change the basis from free to dressed partons) by introducing the coherent operator 
\begin{align}\label{c}
    C = {\cal P} e^{i  \sqrt{2} \int d^2 x  d\xi\, 
    \hat b_a^i(\xi, \v{x}) 
    \left[ 
    a^\dagger_{i,a} (\xi, \v{x})
    + a_{i,a} (\xi, \v{x})
    \right]
    }
\end{align}
where the WW field operator is defined through the total charge density operator $\hat{\rho}_{\rm P}^a(\xi,\v{y})$ integrated from the rapidity of the projectile to $\xi$
\begin{align}
\hat{b}^i_a(\xi,\v{x})=\frac{g}{2\pi}\int d^2y \frac{(\v{x}-\v{y})^i}{|\v{x}-\v{y}|^2} \hat{\rho}_{\rm P}^a(\xi,\v{y})+...
\end{align}
Here we have  explicitly written out only the leading order in $\alpha_s$ expression. Higher order corrections can be included, but we will not take them into account in our calculations. The expectation value of a 
gluon observable $O(a,a^\dagger)$ in the final state is then given by
$ \langle \Psi_{\rm out}  | C  \,O(a,a^\dagger) \, C^\dagger | \Psi_{\rm out}  \rangle  $.

For example the differential single inclusive gluon production is given by 
\begin{align}
    \frac{ dN } { dy d^2 \v q}   =  \langle \Psi_{\rm out}  | C  \, 
    a_{i,b}^\dagger(y, \v q)
    a_{i,b}(y, \v q) \, 
    C^\dagger | \Psi_{\rm out}  \rangle  
\end{align}
This is nothing but the high energy counterpart of the Faddeev-Kulish construction  of dressed states in theories with massless gauge bosons. 

In this paper we will consider double gluon production, which is a straightforward generalization of the previous equation 
\begin{align}
    \frac{dN}{d\eta d^2\v{q}_1 d\xi d^2\v{q}_2} =  \langle \Psi_{\rm out}  | C  \, 
    a_{i,b}^\dagger(\eta, \v{q}_1)
    a_{i,b}^\dagger(\xi, \v{q}_2)
    a_{i,b}(\eta, \v{q}_1) \, 
    a_{i,b}(\xi, \v{q}_2) \, 
    C^\dagger | \Psi_{\rm out}  \rangle  
\end{align}

To perform the actual calculations, we need of course a model for the wave function of the projectile. In general, in the high energy CGC approach, 
this wave function has the form of a wave function for valence charges dressed by the WW field, that is \begin{align}
    | \Psi_{\rm in} \rangle  = C | v \rangle,  
\end{align}
where $C$ is the coherent operator Eq.(\ref{c}) and $| v \rangle$ describes a state in the free Fock space of a small number of valence (large rapidity) partons. As mentioned in the introduction, we use two models for the valence wave function $|v \rangle $ - the dipole and MV model, which we describe in detail below. 

Finally, the target is treated as an ensemble of classical gluon field configurations which have to be averaged over. 

We now proceed with the detailed calculations within the framework just described. 

\subsection{Modelling the Projectile and the Target} 
As stated earlier, we explore two models for the wave function of a nearly on shell photon. 

\subsection{Dilute dipole projectile}
For a nearly on-shell photon, the longitudinal polarization is suppressed by a square of the virtuality.  Hence only the transverse polarization is of importance. 
The leading order perturbative expression for the photon state is given by ~\cite {Beuf:2016wdz}  
\begin{equation}
\begin{split}
|\gamma_{\lambda}^*\rangle \simeq & 
\sum_{s_1,s_2}\int_0^{\infty} \frac{dz}{4(2\pi) z(1-z)} \int \frac{d^2 \v{k}_1 }{(2\pi)^2} \Psi^{T}_{\lambda}(z, \v{k}_1,s_1,s_2)b^{\dagger}_{\alpha,s_1}(\v{k}_1^+, k_{1})d^{\dagger}_{\alpha,s_2}(\v{k}_2^+, -\v{k}_{1})|0\rangle\\
=& \sum_{s_1,s_2}\int_0^{\infty} \frac{dz}{4(2\pi) z(1-z)} \int d^2 \bm{z}_1 d^2 \bm{z}_2 \Psi^{T}_{\lambda}(z, \v r,s_1,s_2)b^{\dagger}_{\alpha,s_1}(\v{k}_1^+, \v{z_1})d^{\dagger}_{\alpha,s_2}(\v{k}_2^+, \v{z_2})|0\rangle\\
\end{split}
\end{equation}
where $s_{1,2}$ are polarizations of quark and anti-quark, $\alpha$ is the color index in the fundamental representation, $\lambda=\pm 1$ is the photon polarization, $r=z_1-z_2$.  The longitudinal momentum of the (anti)quark is ${k}_1^+ =  z p^+$ (${k}_2^+ =  (1-z) p^+$).  The above expression is written in the reference frame where the photon has zero transverse momentum.  The photon splitting functions in the momentum and coordinate space are 
\begin{align}
\Psi^{T}_{\lambda}(z,\v k_{1},s_1,s_2)&=-2ee_f\delta_{s_1,-s_2}(2z-1+2\lambda s_1)\sqrt{z(1-z)}\frac{\v k_{1}\cdot \v{\epsilon_{\lambda}}}{\v k_{1}^2+\varepsilon_f^2}
\\
\Psi^{T}_{\lambda}(z, \v r ,s_1,s_2)&=-i\frac{2ee_f}{2\pi}\delta_{s_1,-s_2}(2z-1+2\lambda s_1)\sqrt{z(1-z)}\frac{\v r \cdot \v{\epsilon_{\lambda} }}{|\v r|}\varepsilon_fK_1(\varepsilon_f|\v r|), 
\end{align}
{%\color{red} 
In the perturbative photon wave function one has $\varepsilon_f^2  = z(1-z)Q^2$. This $z$ dependence of $\varepsilon$ is such that for $z=0$ and $z=1$ it vanishes, and therefore for these ``alligned jet'' configurations the values of the dipole size $r$ can grow arbitrarily large. For our goal,  a  model of an almost real photon with transverse size determined by a nonperturbative soft scale is required. To achieve this, we will modify the perturbative wave function in the simplest possible way, i.e. will take $\varepsilon=Q/2$ independent of the value of $z$.}

In numerical calculations we use a small but nonzero value of $Q\sim 200$ MeV, see Sec.~\ref{Sec:Num} for details. 

\subsubsection{Correlators of the charge density.}
For the calculation we will need to know correlators of up to four color charge density operators. Here we calculate those in the dipole model of the photon.
The charge density operator in terms of the quark and anti quark creation and annihilation operators is defined as
\begin{align}
\hat{\rho}^a(x^-,\v x)= b^{\dagger}_{\alpha \sigma}(x^-, \v{x}) t^a_{\alpha \beta} b_{\beta \sigma}(x^-,\v{x})-d^{\dagger}_{\alpha \sigma}(x^-, \v{x}) t^a_{\beta \alpha } d_{\beta \sigma}(x^-,\v{x})
\end{align}

For the purpose of a CGC calculation we need the correlators of the integrated quantity
\begin{equation}
    \hat\rho^a(\v x)\equiv \int dx^-\hat{\rho}^a(x^-,\v x)
\end{equation}
We now evaluate the correlators of $\hat\rho$ in a dipole state  
 $|q\bar{q}\rangle = b^{\dagger}_{\alpha \sigma}(\v{z_1})d^{\dagger}_{\alpha \sigma}(\v{z_2})|0\rangle$, 
 where we have suppressed indices and arguments irrelevant for this computation. \\[10pt]
 Using 
\begin{equation}
\begin{split}
&\hat{\rho}^a(\v x) b^{\dagger}_{h}(\v {z_1})d^{\dagger}_{h}(\v{z_2})|0\rangle\\
=& \left[ t^a_{\alpha h}b^{\dagger}_{\alpha}(\v{z_1})d^{\dagger}_{h}(\v{z_2}) \delta^{(2)}(\v{x}-\v{z_1})
-t^a_{ h \alpha}b^{\dagger}_{h}(\v{z_1})d^{\dagger}_{\alpha}(\v{z_2})\delta^{(2)}(\v{x}-\v{z_2})\right]|0\rangle
\end{split}
\end{equation}
and 
\begin{equation}
\begin{split}
& \langle 0|d_{l}(\v{z_2})b_{l}(\v{z_1})\hat{\rho}^a(\v{x})\\
=&\langle 0| \left[ t^a_{l \beta}d_{\beta}(\v{z_2})b_{l}(\v{z_1}) \delta^{(2)}(\v{x}-\v{z_1})-t^a_{ \beta l}d_{\beta}(\v{z_2})b_{l}(\v{z_1})\delta^{(2)}(\v{x}-\v{z_2})\right]
\end{split}
\end{equation}
we obtain 
\begin{equation}
\langle q\bar{q} | \hat \rho^{a}(\v{x_1})  \hat\rho^{b}(\v{x_2})  |  q\bar{q}\rangle  
=\frac{\delta^{ab}}{2}
\prod_{i=1,2}
\left( \delta^{(2)}(\v{x_i}-\v{z_1})   - \delta^{(2)}(\v{x_i}-\v{z_2}) \right)\,. 
\end{equation}
For higher correlators, to simplify expressions, we account explicitly for the symmetry of the photon splitting function 
$\v{z_1} \leftrightarrow \v{z_2}$ (and $z \leftrightarrow 1 -z$)\footnote{That is we 
consider  $|q\bar{q}\rangle = \frac12 \left[ b^{\dagger}_{\alpha \sigma}(\v{z_1},z)d^{\dagger}_{\alpha \sigma}(\v{z_2},1-z) + 
b^{\dagger}_{\alpha \sigma}(\v{z_2},1-z)d^{\dagger}_{\alpha \sigma}(\v{z_1},z)\right]
|0\rangle$
}
to obtain 
\begin{equation}
    \begin{split}
        & \langle q\bar{q} |  \hat\rho^{a}(\v{x_1})  \hat\rho^{b}(\v{x_2})  \hat\rho^{c}(\v{x_3})  |  q\bar{q}\rangle   \\
      =& \frac{if_{abc}}{4}  \left( \delta^{(2)}(\v{x_2}-\v{z_1})  + \delta^{(2)}(\v{x_2}-\v{z_2}) \right)  \prod_{i=1,3} \left( \delta^{(2)}(\v{x_i}-\v{z_1})  - \delta^{(2)}(\v{x_i}-\v{z_2}) \right)  
    \end{split}
\end{equation}
and 
\begin{equation}
    \begin{split}
     &   \langle q\bar q|  \hat\rho^{a}(\v{x_1})  \hat\rho^{b}(\v{x_2})  \hat\rho^{c}(\v{x_3}) \hat\rho^{d}(\v{x_4})|q\bar q\rangle 
     =  \Tr(t^at^bt^ct^d)\prod_{i=1}^4\left[ \delta^{(2)}(\v{x_i}-\v{z_1})   - \delta^{(2)}(\v{x_i}-\v{z_2}) \right]\\   
     -&\frac{f_{cde}f_{abe}}{8} \left[\delta^{(2)}(\v{x_2}-\v{z_1})+\delta^{(2)}(\v{x_2}-\v{z_2})\right]   \left[\delta^{(2)}(\v{x_3}-\v{z_1}) +  \delta^{(2)}(\v{x_3}-\v{z_2})\right]
     \\ &\quad  \quad  \quad \times
     \prod_{i=1,4}\left[ \delta^{(2)}(\v{x_i}-\v{z_1})   - \delta^{(2)}(\v{x_i}-\v{z_2}) \right]
    \end{split}
\end{equation}  
where $\Tr(t^a t^b t^c t^d)=\frac{1}{4N_c}\delta^{ab}\delta^{cd}+\frac{1}{8}\left(d^{abe}+if^{abe}\right)\left(d^{cde}+if^{cde}\right)$. 

In the following we will need a representation of this trace in terms its fully symmetric and the remaining parts:
 $\Tr(t^a t^b t^c t^d) = \Tr(t^a t^b t^c t^d)_{\rm sym}
 + \Tr(t^a t^b t^c t^d)_{\rm ns}$
 with 
\begin{equation}
    \begin{split}
        \Tr(t^a t^b t^c t^d)_{\rm sym}&=\frac{1}{12}\Big[\frac{1}{N_c}(\delta^{ab}\delta^{cd}+\delta^{ac}\delta^{bd}+\delta^{ad}\delta^{bc})+\frac{1}{2}(d_{abe}d_{cde}+d_{ace}d_{bde}+d_{ade}d_{bce})       \Big]
    \end{split}
\end{equation}
and 
\begin{equation}
    \begin{split}
        \Tr(t^a t^b t^c t^d)_{\rm ns}&=
        \frac{1}{12} (f_{ade} f_{bce} - f_{abe} f_{cde})
        + \frac{1}{8} i (d_{abe} f_{cde} + d_{cde} f_{abe} )\,.
    \end{split}
\end{equation}
For $N_c=3$, one can  use the following identity (the factor of two was missing in Refs.~\cite{haber2016useful}; the validity of our identity can be easily checked by summing with respect to $a=b$) 
\begin{align}
    d_{abe}d_{cde}+d_{ace}d_{bde}+d_{ade}d_{bce}=\frac{2}{3}(\delta^{ab}\delta^{cd}+\delta^{ac}\delta^{bd}+\delta^{ad}\delta^{bc})
\end{align}
to obtain 
\begin{equation}
\label{Eq:TrSym}
    \begin{split}
        \Tr(t^a t^b t^c t^d)^{N_c=3}_{\rm sym}
        &=\frac{1}{18}          (\delta^{ab}\delta^{cd}+\delta^{ac}\delta^{bd}+\delta^{ad}\delta^{bc})\,.
    \end{split}
\end{equation}
We stress again  that this expression is only true for $N_c=3$.

\subsection{The MV model for the projectile}
The other model we use to describe the photon wave function is the McLerran-Venugopalan model. It 
posits that the averaging over the ``classical'' color charge density of the projectile should be performed with Gaussian weight, so that all correlators are expressed in terms of Wick contractions of the basic "propagator"
\begin{align}
\label{Eq:rho2mv1}
 \langle \rho_a(\v x) \rho_b(\v y) \rangle_{\rm MV} 
 =   \mu^2(x) \delta^{(2)}(\v x- \v y) \delta_{ab}\,.
 \end{align}
  Note that \eqref{Eq:rho2mv1} goes beyond the conventional MV model in that it includes the dependence of the color charge density on the impact parameter $\v{b} = \frac{\v{x}+\v{y}}{2}$. In principle one can go even further and account for the color neutrality by substituting a suitable function instead of  $\delta^{(2)}(\v x- \v y)$, see e.g. Ref.~\cite{McLerran:2016ivs}. We will not do this in the current paper. 
  
 One has to be aware that the ``classical'' charge density in \eqref{Eq:rho2mv1} is not identical to the color charge density operator $\hat \rho$. To calculate the correlator of color charge density operators, one has to perform the operator symmetrization first, and use the MV for the completely symmetric parts. This was shown in \cite{PhysRevD.72.074023} and explained from the point of view of quantum-classical correspondence in \cite{Li:2020bys}.
 
 For bilinears, the symmetrization is straightforward:
\begin{align}
 \hat \rho_a(\v x) \hat \rho_b(\v y) 
 &=\frac12 \left\{ \hat \rho_a(\v x),  \hat \rho_b(\v y) \right\} +  \frac12 \left[ \hat \rho_a(\v x),  \hat \rho_b(\v y) \right] \notag \\ 
 &= \rho_a(\v x) \rho_b(\v y) - \frac{1}{2} \delta^{(2)}(\v x- \v y) T^{c}_{ab} 
 \rho_c (\v x)
\end{align}
where we have replaced the fully symmetric combinations by the classical color density $\left\{ \hat \rho_a(\v x),  \hat \rho_b(\v y) \right\} \to \rho_a(\v x) \rho_b(\v y)$ and took into account that the commutator $\left[ \hat \rho_a(\v x),  \hat \rho_b(\v y) \right] = - \frac{1}{2} \delta^{(2)}(\v x-\v y) T^{c}_{ab} \rho_c (\v x) $ as follows from the definition of $\hat \rho$. We thus see that 
\begin{align}
\label{Eq:rho2mv}
 \langle \hat \rho_a(\v x) \hat \rho_b(\v y) \rangle_{\rm MV} 
  =  \langle \rho_a(\v x) \rho_b(\v y) \rangle_{\rm MV} 
 =   \mu^2(x) \delta^{(2)}(\v x- \v y) \delta_{ab}\,.
 \end{align}
This procedure can be extended to any number of  operators $\hat \rho$.  

For three operators we obtain: 
\begin{align}
    %\langle 
    \notag
    &\hat \rho_a(\v x) \hat \rho_b(\v y) \hat 
    \rho_c(\v z) 
    =  \rho_a(\v x) \rho_b(\v y) 
    \rho_c(\v z)  \\ &- 
    \frac12 \big( \DDelta{\v y}{\v z} T^e_{ b c} \rho_a(\v x) \rho_e (\v y) +
    \DDelta{\v x}{\v z} T^e_{ a c} \rho_b (\v y) \rho_e(\v x) +
   \DDelta{\v x}{\v y} T^e_{a b} \rho_e(\v  x) \rho_c(\v z)\big)\,,
    %\rangle_{\rm MV}  \to   
\end{align}
 with $T^{a}_{bc} = - i f_{abc}$.
Here the terms linear in $\hat \rho$ were omitted as they do not contribute to the expectation value over the color invariant ensemble~\footnote{Note that this expression also reproduces the result of the previous section. Indeed, taking into account the symmetry of the dipole wave function we need to account only for the term quadratic in $\rho$
\begin{align}
    \langle \rho_a(\v x)  \rho_b(\v y) \rangle 
     = \langle  \{ \rho_a(\v x)  , \rho_b(\v y) \} \rangle
     = \frac{\delta_{ab}}{ 2 }  (\delta(\v x-\v {z_1})-\delta(\v{x}-\v{z_2}))(\delta(\v{y}-\v{z_1})-\delta(\v{y}-\v{z_2})) \,.
     \notag
\end{align}
Therefore 
\begin{align}
    \langle \rho_a(\v x)  \rho_b(\v y) \rho_c(\v z) \rangle 
 = &- 
    \frac1{4} T^{a}_{bc} 
    \Big[ \DDelta{\v y}{\v z} (\delta(\v x-\v{z_1})-\delta(\v{x}-\v{z_2}))(\delta(\v{y}-\v{z_1})-\delta(\v{y}-\v{z_2}))
   \notag \\ & -
    \DDelta{\v x}{\v z}  
    (\delta(\v {x} -\v{z_1})-\delta(\v{x}-\v{z_2}))(\delta(\v{y}-\v{z_1})-\delta(\v{y}-\v{z_2})) 
    \notag \\ & + 
   \DDelta{\v x}{\v y} (\delta(\v x- \v{z_1})-\delta(\v x-\v{z_2}))(\delta(\v{z}-\v{z_1})-\delta(\v{z}-\v{z_2}))  \Big] \notag \\
 & =  - 
    \frac1 {4} T^{a}_{bc} 
    (\delta(\v{x}-\v{z_1})-\delta(\v{x}-\v{z_2}))
    (\delta(\v{z}-\v{z_1})-\delta(\v{z}-\v{z_2}))
    (\delta(\v{y}-\v{z_1})+\delta(\v{y}-\v{z_2}))\,.
\notag
\end{align} 
}.
Thus in the MV model : 
\begin{align}
\label{Eq:rho3mv}
    \langle 
    \hat \rho_a(\v x) \hat \rho_b(\v y) \hat 
    \rho_c(\v z)
    \rangle_{\rm MV}
    &=  
     -   \frac12  \DDelta{\v x}{\v y} \DDelta {\v y} {\v z} T^a_{bc} \, \mu^2 (\v x)
    %  \to   
\end{align}
Similarly for four operators, we obtain 
\begin{align}
\notag 
    &\langle 
    \hat \rho_a(\v x) \hat \rho_b(\v y) \hat 
    \rho_c(\v z) \hat\rho_d(\v u)
    \rangle_{\rm MV}
    \\ &   =  
    \mu^2(\v x) \mu^2(\v z) \delta^{ab}\delta^{cd} \DDelta{\v x}{\v y}  \DDelta{\v z}{\v u}
    + \mu^2(\v x) \mu^2(\v y) \delta^{ac}\delta^{bd}  \DDelta{\v x}{\v z}  \DDelta{\v y}{\v u} 
        \notag \\ & 
     + \mu^2(\v x) \mu^2(\v z)  \delta^{ad}\delta^{bc} \DDelta{\v x}{\v u}  \DDelta{\v z}{\v y} 
         \notag \\ &   + 
     \frac{\mu^2(x)}{6} \DDelta{\v x}{\v y} \DDelta {\v x} {\v z}  \DDelta {\v x} {\v u}   \left( 
          f_{a d e}  f_{b c e} -  f_{a b e} f_{c d e}  
     \right) 
     \label{Eq:rho4mv}
    %  \to   
\end{align}
The last term here is of lower order in $\mu^2$ and is thus subleading for a projectile with  high charge density. We do not neglect it here, but as will become clear later, it does play a somewhat different role than the first two terms even for a dilute projectile.

\subsection{Small $x$ gluon component of the projectile wave function}
\label{Sec:Coherent}
At the leading order, the dressing of the color charge by gluons of rapidity $\eta$ is given by the coherent operator: 
\begin{align}
C_{\eta}=\exp{i\sqrt{2} \int d^2 \v x  \hat{b}^i_a(\v x)\left[ a^{i\dagger}_a(\eta, \v x)+a^i_a(\eta, \v x) \right]}
\end{align}
where the creation/annihilation operators are introduced as the decomposition of the transverse 
gluon field  
\begin{equation}
\begin{split}
A^i_a(x^+,\v x)=\int_{0}^{\infty}\frac{dk^+}{2k^+(2\pi)}\left(a^i_a(k^+,\v x)e^{-ik^-x^+}  +a^{i\dagger}_a(k^+,\v x)e^{ik^-x^+} \right)\,.
\end{split}
\end{equation}
The rapidity variable is defined here as 
\begin{align}
d\eta=\frac{dk^+}{k^+}
\end{align}
and the corresponding 
\begin{align}
a^i_a(\eta, \v x)={\sqrt{\frac{k^+}{4\pi}}} a^i_a(k^+, \v x). 
\end{align}
The commutation relations for the creation/annihilation operators are 
\begin{align}
\left[a^i_a(\eta, \v x), a^{j\dagger}_b(\eta', \v y)\right]=\delta^{ij}\delta^{ab} \delta^2(\eta-\eta') \delta(\v  x - \v y)\,. 
\end{align}
The Weizs\"acker-Williams field at the leading order is proportional to the color charge density operator 
\begin{align}
\hat{b}^i_a(\v x)=\frac{g}{2\pi}\int d^2\v y \frac{(\v x-\v y)^i}{(\v x- \v y)^2} \hat{\rho}^a(\v y)\, . 
\end{align}
The distribution (wave function) of $\hat \rho$ can be modelled by the previously discussed dipole or MV models.

Our calculation requires the knowledge of the incoming wave function at the rapidities of both observed gluons: $\eta$ and $\xi<\eta$. As we alluded to in the introduction, we restrict our calculation  to the case when the difference between the rapidites of the observed gluons $\eta-\xi$ is large enough to use the eikonal vertex, that is $\eta-\xi \gg 1$,  but small enough to neglect  corrections due to evolution, that is $\eta-\xi \ll 1/\alpha_s$.

The gluon with rapidity $\eta$ itself contributes to the source for emission of gluons at lower rapidity.  This can be accounted for by an appropriate redefinition of the Weizs\"acker-Williams field at lower rapidity $\hat{b}^i_a(\v x) \to \hat{b}^i_a(\v x) + \delta \hat{b}^i_a(\v x)$. 
  Thus at rapidity $\xi$ we have
\begin{align}
C_{\xi}=\exp{i\sqrt{2} \int d^2\v x  \left(\hat{b}^i_a(\v x)+\delta \hat{b}^i_a(\v x)\right)\left[ a^{i\dagger}_a(\xi, \v x)+a^i_a(\xi, \v x) \right]}
\end{align}
where the extra contribution due to the presence of the gluon in the dipole wave function is given by 
\begin{equation}
\begin{split}
\delta \hat{b}^i_a(\v x)&=\frac{g}{2\pi} \int_{\xi}^{\eta} d\zeta   \int d^2y \frac{(\v x-\v y)^i}{(\v x- \v y)^2} \hat{\rho}_g^a(\zeta, \v y)
\end{split}
\end{equation}
where 
\begin{align}
\hat{\rho}^a_g(\zeta, \v x)= a^{i\dagger}_b(\zeta, \v x) T^a_{bc} a_c(\zeta, \v x)\,.
\end{align}
%Here we introduced $T^{a}_{bc} = - i f_{abc}$. 

Thus to the required order, the valence state dressed by the gluon cloud is 
\begin{equation} |\Psi_{in}\rangle=C_{\xi}C_{\eta} |0\rangle |v\rangle
\end{equation}

\subsection{Scattering off the dense target}
\label{Sec:Scattering} 

To describe a large nucleus, we use the classic field  approximation for the gluon field of the target. In the eikonal approximation  we have,
\begin{align}
\label{Eq:S_g}
\hat{S}^{\dagger}a^i_{a}(\zeta, \v{\bar{u}})\hat{S}&=U^{aa'}_{\v{\bar{u}}}a^i_{a'}(\zeta, \v{\bar{u}})
\end{align}
for gluon scattering, where $U$ is the infinite Wilson line in the adjoint representation: 
\begin{equation}
        U_{\v{x}}={\cal P}\exp{ig\int_{-\infty}^\infty dx^{+}T^aA_a^-(x^+,\v{x})}\,. 
\end{equation}
Similarly for quark and antiquark 
\begin{equation}
\begin{split}
\hat{S}^{\dagger}b_{\alpha}(\v{x})\hat{S}&=V_{\alpha \beta}(\v{x})b_{\beta}(\v{x})\\
\hat{S}^{\dagger}d_{\alpha}(\v{x})\hat{S}&=V^{\dagger}_{\beta \alpha }(\v{x})d_{\beta}(\v{x})
\end{split}
\end{equation}
where the Wilson line in the fundamental representation is given by 
\begin{equation}
        V(\v{x})={\cal P}\exp{ig\int_{-\infty}^\infty dx^{+}t^aA_a^-(x^+,\v{x})}\,. 
\end{equation}
The consequence of these equations and the definition of the color charge density is that  
\begin{align}
\hat{S}^{\dagger}\hat{\rho}_a(\v{\bar{u}})\hat{S}&=U^{aa'}_{\v{\bar{u}}}\hat{\rho}_{a'}(\v{\bar{u}})
\end{align}
and similarly  
\begin{equation}
\hat{S}^{\dagger}\hat{\rho}_a(\zeta,\v{ \bar{u}})\hat{S}=a^{i\dagger}_{b}(\zeta, \v{\bar{u}}) a_{c}(\zeta, \v{\bar{u}})\left[U^{\dagger}_{\v{\bar{u}}}  T^a  U_{\v{\bar{u}}} \right]_{bc}
 = U^{aa'}_{\v{\bar{u}}} 
 \hat{\rho}_{a'}(\zeta,\v{ \bar{u}})
\,.
\end{equation}

A calculation of a physical observable requires averaging over the ensemble of target fields which determines various correlators of the Wilson lines.
 For a large nucleus at small x, the ensemble of the color fields of the target is frequently taken as an MV model with a relatively large value of the saturation momentum. This procedure in practice works well for calculation of a correlator of a small number of Wilson lines, but becomes increasingly complicated and cumbersome for more complicated correlators.
 
 As we will see, the observable we work with leads to the expectations values of four Wilson lines in the adjoint representation. In the limit of  large number of colors, this can be reduced to a product of up to two dipoles and a quadrupole in the fundamental representation in the coordinate space. 
In order to get particle production cross section, this combination has to be Fourier-transformed with respect to four two-dimensional transverse momenta.  Although, there are analytical expressions for the dipole and quadrupole in the McLerran-Venugopalan model, the Fourier transforms render the problem numerically prohibitively complex, as fast Fourier transform is not feasible in this multidimensional space while Monte-Carlo integration is extremely ineffective due to the presence of the oscillatory phases. 

The two particle correlations at large $N_c$ (in the classical approximation, see e.g. Ref.~\cite{Kovchegov:2018jun})
only arise at order $1/N_c^2$~\cite{Kovner:2010xk,Kovchegov:2012nd,Kovchegov:2013ewa}. Partly this can be attributed to the nature of the correlations arising either from gluon HBT or gluon Bose-Einstein correlations~\cite{Altinoluk:2018ogz}. Therefore the large $N_c$ limit does not capture the physics we want to study.   
Going beyond the large $N_c$ limit provides additional level of complication, as the cross section also involves sixtupoles and octupoles for which explicit analytical expressions, if exist, are too complex for a numerical simulation. 

Instead we will use another approach applicable for a well evolved, near black disc, target. The approximation in question is the so-called {\it factorized  dipole approximation} (FDA) discussed in Ref.~\cite{Kovner:2017ssr,Kovner:2018vec} and successfully applied to study particle production and correlations in Refs.~\cite{Altinoluk:2018ogz,Kovner:2018fxj,Kovner:2018azs,Li:2021ntt,Altinoluk:2020psk}. 

The idea behind this approximation is extremely intuitive. Consider an arbitrary combination of even number of  Wilson lines
$\langle U_{x_1} U_{x_2} \ldots U_{x_{2n}} \rangle$ which is multiplied by  a non-trivial function of coordinates  and integrated with respect to all two-dimensional vectors $x_i$, $i=1, \ldots, 2n$ over some finite, but large compared to the inverse saturation scale of the target squared, transverse area $S_\perp$. If a single Wilson line were to have a nonzero expectation value the largest contribution to the integral would be of order $S^{2n}_\perp$ arising from the integration over the part of the phase space where the coordinates of all Wilson lines are far away from each other. However for a dense target $\langle U_{ab} \rangle = 0$, and so the largest power of the transverse area possible is $S^{n}_\perp$, with corrections of order $S^{n-1}_\perp$.
The leading order contribution includes terms which contain smallest color singlets in the projectile propagating through the target. On one hand, any non singlet state that in the transverse plane separated by more than $1/Q_s$ from other propagating partons must have a vanishing S-matrix on the dense target due to color neutralization on the scale of $1/Q_s$, see e.g. \cite{Iancu:2001md,Ferreiro:2002kv,Iancu:2002xk,Mueller:2002pi}. On the other hand, if the singlet state contains more than two partons, one looses a power of the area when integrating over the coordinates of the partons. Thus the leading contribution in the dense target regime  is the one where only dipole contribution to the S-matrix should be accounted for.
For illustration lets consider a four Wilson line observable   
\begin{align}
    &\int \prod_{i=1}^{4} d^2 \v x_i  f(\v x_1, \ldots , \v x_4)
    \langle {\rm tr}  [U_{\v x_1} U_{\v x_2} U_{\v x_3}  U_{\v x_4}  ]\rangle 
    \notag \\ &=  
    \int \prod_{i=1}^{4} d^2 \v x_i  f(\v x_1, \ldots ,\v x_4)\Big[
    \langle U^{ab}_{\v x_1} U^{bc}_{\v x_2} \rangle 
     \langle U^{cd}_{\v x_3} U^{da}_{\v x_4} \rangle
    \notag \\ & + 
    \langle U^{da}_{\v x_4} U^{ab}_{\v x_1}\rangle 
     \langle  U^{bc}_{\v x_2} U^{cd}_{\v x_3}\rangle
    +
    \langle U^{ab}_{\v x_1} U^{cd}_{\v x_3}  \rangle 
     \langle  U^{bc}_{\v x_2} U^{da}_{\v x_4} \rangle
     \Big]
      + {\cal O}(S_\perp) \notag 
\end{align}
where the function  $f$ is assumed to have support for all $x$ within the area $S_\perp$. The correlator   $\langle U^{ab}_{\v x} U^{cd}_{\v y} \rangle$ is non zero only  for a color singlet state, in this case -- the  dipole: 
\begin{align}
    \label{Eq:KovnersHammer}
    \langle U^{ab}_{\v x} U^{cd}_{\v y} \rangle = \frac{1}{N_c^2-1} \delta_{ac} \delta_{bd} D(|\v x-\v y|)\,, 
\end{align}
where $D(r) = \frac{1}{N_c^2-1} {\rm tr} [ U^{\dagger}_{\v r} U_{0} ]$. 
The quality of this approximation was explicitly checked for a number of Wilson line combinations in Ref.~\cite{Li:2021ntt}. The key necessary (albeit not sufficient) condition for its applicability is  $Q^2_s S_{{\rm proj}, \perp} \gg 1$, where $Q_s$ is the saturation momentum of the nucleus and $S_{{\rm proj}, \perp}$ is the effective transverse area of the projectile. For LHC energies 
this condition is satisfied if we use the natural $ S_{{\rm proj}, \perp} = \frac{1}{Q_{\rm eff}^2}$ where $Q_{\rm eff}$ is the larger between  the QCD non-perturbative scale $\Lambda_{\rm QCD}$ and the virtuality of the photon.  For the nearly real photon we thus have $S_{{\rm proj}, \perp} \sim \Lambda_{\rm QCD}^{-2}$. One has to be aware that when  large momentum particle production is considered the effective area of the projectile might be significantly smaller than  $\Lambda_{\rm QCD}^{-2}$. Although this is a process-dependent statement, in general it means that at  large momenta $k\gg Q_s$, the FDA is not uniformly applicable. {%\color{red}
Nevertheless even for high momenta the approximation can be justified {\it a posteriori} for many momentum configurations.
In particular, using the MV model for near real photon, we checked the quality of the FDA for contribution to double gluon production dominant at large density, see Eq.~\eqref{Eq:54}, and found excellent agreement.  %The classical contribution involved only a limited set of Wilson line correlator and cannot justify the applicability of the approximation in general. 
A somewhat different set of  Wilson line correlators were considered in Ref.~\cite{Li:2021ntt}. Most of the combinations show excellent agreement. At the same time, Ref.~~\cite{Li:2021ntt}, also identified  a few examples where the FDA fails.  These particular correlators do not appear in our expressions required to compute azimuthal anisotropy.}

For calculations in this paper we apply the FDA to evaluate the correlators of more than
two  Wilson lines. As our calculations are mostly in  momentum space, the basic correlator used in the FDA  is the Fourier transform of   Eq.~\eqref{Eq:KovnersHammer}:
\begin{align}
    \langle U^{ab}_{\v{p}} U^{cd}_{\v{q}} \rangle_T=\frac{(2\pi)^2}{N_c^2-1}\delta_{ac}\delta_{bd}\delta^{(2)}( \v{p}+ \v{q})D(\v{p})
\end{align}
where $D(\v{p})$ is
\begin{align}
    D(\v{p})&=\frac{1}{(N^2_c-1) S_{\perp}}\langle \Tr( U^{\dagger}_{\v{p}}U_{\v{p}})\rangle_T\,.
\end{align}
We will also assume spatial isotropy of the target field ensemble, so that  $D( \v{p}) = D(|\v p|)$. 
We will use the MV model for the dipole. That is in the coordinate space, we have 
\begin{align}
    D(\v r) = \exp\left[-\frac{1}{4}Q_s^2r^2\ln(\frac{1}{\Lambda^2r^2}) \right]\, 
\end{align}
with $D(|\v p|) = \int d^2 r e^{i \v p\cdot \v r}    D(\v r)$.

\section{Double inclusive gluon production - general expressions}\label{sec3}

The differential cross section for double inclusive gluon production with  rapidities and the transverse momenta $(\eta, \v{q}_1)$ and $(\xi, \v{q}_2)$ is given by 
\begin{equation}
\label{Eq:Double}
\begin{split}
\frac{d{\cal N}}{d\eta d \v{q}_1^2 d\xi d \v{q}_2^2}=&\frac{1}{(2\pi)^4} \int
d^2 \v u_1 d^2\v u_2 
d^2  \bar{\v u}_1 d^2  \bar{\v u}_2 
e^{-i\v{q}_1(\v{u}_1-\v{\bar{u}}_1)}e^{-i\v{q}_2(\v{u}_2-\v{\bar{u}}_2)}
\\ &\times
\langle \gamma^*|C^{\dagger}\hat{S}^{\dagger}C a_{i,a}^{\dagger}(\eta, \v{u}_1)a_{j,b}^{\dagger}(\xi, \v{u}_2)  
a_{i,a}(\eta, \v{\bar{u}}_1)a_{j,b}(\xi, \v{\bar{u}}_2)C^{\dagger}\hat{S}C| \gamma^* \rangle
\end{split}
\end{equation} 
The coherent operator $C$ is given by the product of $C = C_\xi C_\eta$, see Sec.~\ref{Sec:Coherent}. For double-inclusive gluon production, it is sufficient to expand the coherent operator to the leading order in the argument:
\begin{equation}
\begin{split}
C_{\eta}\simeq&      1+i\sqrt{2} \int d^2\v v_1 \hat{b}^i_{a}(\v{v}_1)\left[ a^{i\dagger}_a(\eta, \v{v}_1)+a^i_a(\eta, \v{v}_1) \right]\,,\\
C_{\xi}\simeq&  1+i\sqrt{2} \int d^2\v v_2 \left(\hat{b}^j_{b}(\v{v}_2)+\delta \hat{b}^j_b(\eta, \v{v}_2)\right)\left[ a^{j\dagger}_b(\xi, \v{v}_2)+a^j_b(\xi, \v{v}_2)\right] \,. 
\end{split}
\end{equation}
Hence to the required order, the coherent operator can be approximated by $C_{\rm eff}$
\begin{equation}
\begin{split}
C_{\rm eff} =&1+i\sqrt{2} \int d^2\v v_1 \hat{b}^i_{a}(\v v_1)\left[ a^{i\dagger}_a(\eta, \v{v}_1)+a^i_a(\eta, \v{v}_1) \right]\\
&+i\sqrt{2} \int d^2\v v_2 \left(\hat{b}^j_{b}(\v{v}_2)+\delta \hat{b}^j_b(\eta, \v{v}_2)\right)\left[ a^{j\dagger}_b(\xi, \v{v}_2)+a^j_b(\xi, \v{v}_2)\right] \\
&-2 \int d^2\v v_1d^2\v v_2 \left(\hat{b}^j_{b}(\v{v}_2)+\delta \hat{b}^j_b(\eta, \v{v}_2)\right)\left[ a^{j\dagger}_b(\xi, \v{v}_2)+a^j_b(\xi, \v{v}_2)\right] \notag \\ &\quad \quad  \quad \quad  \quad \quad  \times  \hat{b}^i_{a}(\v{v}_1)\left[ a^{i\dagger}_a(\eta, \v{v}_1)+a^i_a(\eta, \v{v}_1) \right]\,. 
\end{split}
\end{equation}
The matrix element in Eq.~\eqref{Eq:Double},  can be decomposed into three different contributions classified by the number of gluons emitted directly from the photon state $\mathcal{M}=\Sigma_2+\Sigma_3+\Sigma_4$, or equivalently, the powers of color charge density, schematically, 
\begin{align}
b^2(b+\delta b)^2=b^4+2b^3 \delta b+ b^2(\delta b)^2\,.
\end{align}
Here every factor of $b$ corresponds to a gluon emitted directly from the photon, while a factor of $\delta b$ corresponds to a gluon emitted from another gluon with larger rapidity.
We will denote the corresponding contribution to the matrix element by $\Sigma_n$, where $n$ is the number of gluons emitted directly from the projectile. To elaborate on this: $\Sigma_2$ contributes to the probability of the process where a gluon with rapidity $\eta$ is emitted from the photon, and subsequently this gluon splits into a pair of gluons which are then observed in the final state.  The splitting can occur either before of after the scattering on the target field. $\Sigma_4$ contributes to the probability of the process where both gluons are emitted directly from the photon state. Finally $\Sigma_3$ represents the interference of the previous two contributions. 

%In terms of the production amplitudes we have 
%\begin{align}
 %   {\cal M} = \left( \langle 1^\prime  | + \langle 2^\prime  |   \right)\left( | 1  \rangle  + | 2  \rangle   \right) 
  %   = 
   %   \langle 1^\prime  |  1  \rangle
    %  + \left( \langle 1^\prime  |  2  \rangle
    %  + \langle 2^\prime  |  1  \rangle \right)+
    %  \langle 2^\prime  |  2  \rangle
%\end{align}
%where $ |  1  \rangle$ ($|  2  \rangle$)
 %denotes an amplitude with one (two) gluons
%emitted directly from the large $x$ wave function of the projectile. They are trivially related to the aforementioned $\Sigma_n$, $\Sigma_2 =  \langle 1^\prime  |  1  \rangle$, $\Sigma_3 =  \langle 1^\prime  |  2  \rangle
 %     + \langle 2^\prime  |  1  \rangle$ and $\Sigma_4 =  \langle 2^\prime  |  2  \rangle$.   
We now turn to evaluating the production amplitude. 

\begin{figure}[t!]
	\centering 
	\includegraphics[width=0.4\textwidth]{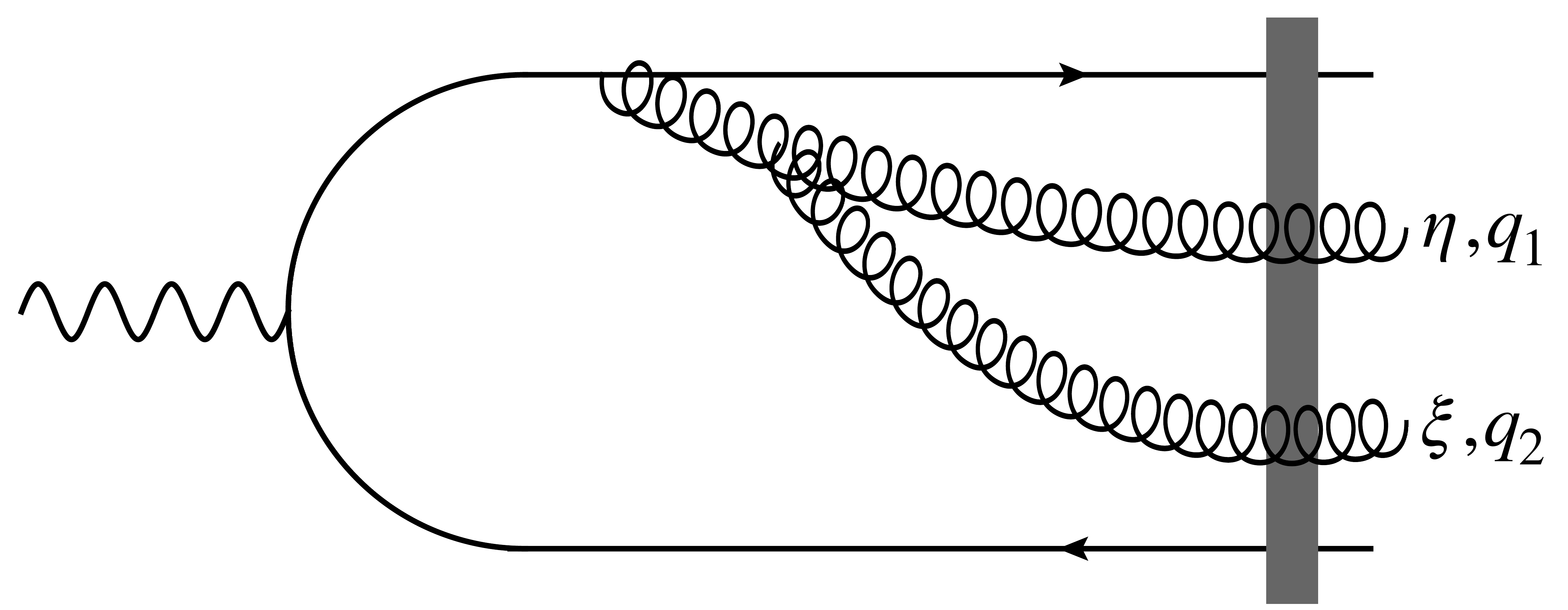} \quad \quad 
 	\includegraphics[width=0.4\textwidth]{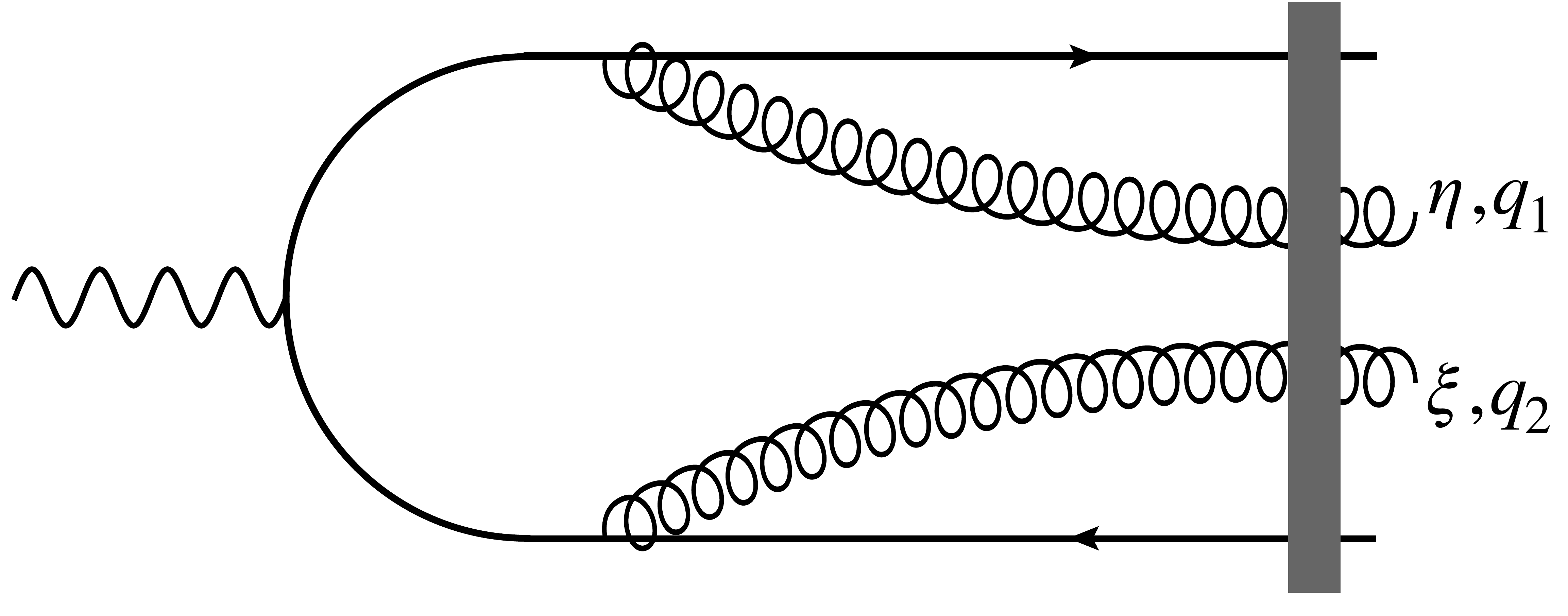}
	\caption{An illustration of two types of contribution to the two gluon production amplitude. Note that the number of actual diagrams is significantly larger, as the position of the shock wave can vary. For example, in total, there are six diagrams for the topology of the left diagram.  }
\end{figure}

\subsection{Production amplitude}
Our goal is to evaluate the production amplitude in the coordinate space defined by 
\begin{align}
    {\cal A}(\bar{\v u}_1,\bar{\v u}_2) = 
\hat S^\dagger  a_{i,a}(\eta, \v{\bar{u}}_1)a_{j,b}(\xi, \v{\bar{u}}_2)C^{\dagger}\hat{S}C| \gamma^* \rangle\,.
\end{align}
We have introduced the factor $\hat S^\dagger$ for convenience, although it cancels in the calculation in the matrix element ${\cal M} = {\cal A}^\dagger {\cal A}$.
Using Eq.~\eqref{Eq:S_g}, we arrive at 
\begin{align}
    {\cal A}(\bar{\v u}_1,\bar{\v u}_2) = 
 U^{aa'}_{\v{\bar{u}}_1}U^{bb'}_{\v{\bar{u}}_2}
  a_{   i,a}(\eta, \v{\bar{u}}_1)a_{j,b}(\xi, \v{\bar{u}}_2)
\hat S^\dagger C^{\dagger}\hat{S}C| \gamma^* \rangle \equiv {\cal A}_1+{\cal A}_2\,.
\end{align}
The operator $\hat S^\dagger C^{\dagger}\hat{S}C$ has terms proportional to $ b$ and to $b^2$, and these terms have been separated in the previous equation into the two amplitudes ${\cal A}_1$ and ${\cal A}_2$.
These amplitudes are directly related to  the aforementioned $\Sigma_n$, $\Sigma_2 = {\cal A}_1{\cal A}^*_1$, $\Sigma_3 = {\cal A}_1{\cal A}^*_2+{\cal A}^*_1{\cal A}_2$  and $\Sigma_4 = {\cal A}_2{\cal A}^*_2$.

\subsubsection{Amplitude ${\cal A}_1$}
Expanding the coherent operators and collecting the terms of first order in $b$, we obtain 
the production amplitude for two gluons with colors $a,b$, polarizations $i,j$ at positions $\v{\bar{u}}_1$ and $\v{\bar{u}}_2$:
\begin{equation}
\begin{split}
&{\cal A}_1(\bar{\v u}_1,\bar{\v u}_2)\\
&= 2\int d^2\v v_1d^2\v v_2 
U^{aa'}_{\v{\bar{u}}_1} U^{bb'}_{\v{\bar{u}}_2}a^i_{a'}(\eta, \v{\bar{u}}_1)
a^j_{b'}(\xi, \v{\bar{u}}_2)
\left[ \hat S^{\dagger} \delta \hat{b}^k_c(\eta, \v{v}_2)a^{k\dagger}_c(\xi, \v{v}_2)\hat S\hat{b}^l_{d}(\v{v}_1)a^{l\dagger}_d(\eta, \v{v}_1) \right.\\
&\left.-  \hat S^{\dagger} \hat{b}^l_{d}(\v{v}_1)a^{l\dagger}_d(\eta, \v{v}_1)\delta \hat{b}^k_c(\eta, \v{v}_2)a^{k\dagger}_c(\xi, \v{v}_2)\hat S- \delta \hat{b}^k_c(\eta, \v{v}_2)a^{k\dagger}_c(\xi, \v{v}_2)\hat{b}^l_{d}(\v{v}_1)a^{l\dagger}_d(\eta, \v{v}_1) \right]|\gamma^*\rangle
\end{split}\label{Eq:am1}
\end{equation}
The second term in this expression gives zero when the operator $\delta \hat{b}^k_c(\eta, \v{v}_2)$ acts on the gluon vacuum at rapidity $\eta$ (we always assume that $\xi < \eta$). 

We thus have two non-trivial terms which can be readily evaluated. The first one gives  
\begin{equation}
\begin{split}
&U^{aa'}_{\v{\bar{u}}_1}U^{bb'}_{\v{\bar{u}}_2}\int d^2\v v_1d^2\v v_2 a^i_{a'}(\eta, \v{\bar{u}}_1)a^j_{b'}(\xi, \v{\bar{u}}_2) \hat S^{\dagger} \delta \hat{b}^k_c(\eta, \v{v}_2)a^{k\dagger}_c(\xi, \v{v}_2)\hat S\hat{b}^l_{d}(\v{v}_1)a^{l\dagger}_d(\eta, \v{v}_1)|\gamma^*\rangle\\
%&=U^{aa'}_{\v{\bar{u}}_1}U^{bb'}_{\v{\bar{u}}_2}\int^{\eta}_{\xi} d\zeta \int d^2 x \int d^2v_1d^2v_2 a^i_{a'}(\eta, \v{\bar{u}}_1)a^j_{b'}(\xi, \v{\bar{u}}_2) f^k(\v{v}_2-\v{x})a^{i'\dagger}_{e}(\zeta, \v{x}) \\&a^{i'}_{f}(\zeta, \v{x})a^{k\dagger}_{d'}(\xi, \v{v}_2)U^{\dagger d'c}(\v{v}_2)\left[U^{\dagger}_{\v{x}}  T^c  U_{\v{x}} \right]_{ef}\hat{b}^l_{d}(\v{v}_1)a^{l\dagger}_d(\eta, \v{v}_1)|0,v\rangle\\&=U^{aa'}_{\v{\bar{u}}_1}U^{bb'}_{\v{\bar{u}}_2} f^j(\v{\bar{u}}_2-\v{\bar{u}}_1)U^{\dagger b'c}(\v{\bar{u}}_2)\left[U^{\dagger}_{\v{\bar{u}}_1}  T^c  U_{\v{\bar{u}}_1} \right]_{a'f}\hat{b}^i_{f}(\v{\bar{u}}_1)|0,v\rangle\\
&= \int d^2 \v x  f^j({\bar{\v u}}_2-{\bar{\v u}}_1)f^i(\v{\bar{u}}_1-\v{x})\left[ T^b  U_{\v{\bar{u}}_1} \right]_{af}\hat\rho_f(\v{x})|\gamma^*\rangle
\end{split}
\end{equation}
where 
\begin{align}
     f^i(\v{x}) = \frac{ g } { 2\pi} \frac{ \v x_i} {\v x^2} \,.
\end{align}
The last term in Eq.~\eqref{Eq:am1} simplifies into 
\begin{equation}
\begin{split}
&U^{aa'}_{\v{\bar{u}}_1}U^{bb'}_{\v{\bar{u}}_2}\int d^2\v v_1d^2\v v_2 a^i_{a'}(\eta, \v{\bar{u}}_1)a^j_{b'}(\xi, \v{\bar{u}}_2) \delta \hat{b}^k_c(\eta, \v{v}_2)a^{k\dagger}_c(\xi, \v{v}_2)\hat{b}^l_{d}(\v{v}_1)a^{l\dagger}_d(\eta, \v{v}_1)|\gamma^*\rangle\\
%=&U^{aa'}_{\v{\bar{u}}_1}U^{bb'}_{\v{\bar{u}}_2}\int d^2v_1 a^i_{a'}(\eta, \v{\bar{u}}_1) \delta \hat{b}^j_{b'}(\eta, \v{\bar{u}}_2)\hat{b}^l_{d}(\v{v}_1)a^{l\dagger}_d(\eta, \v{v}_1)|\gamma^*\rangle\\
%=&U^{aa'}_{\v{\bar{u}}_1}U^{bb'}_{\v{\bar{u}}_2}\int d^2 y \int d^2v_1 a^i_{a'}(\eta, \v{\bar{u}}_1) f^j(\v{\bar{u}}_2-\v{y}) a^{j'\dagger}(\eta, \v{y})T^{b'}a^{j'}(\eta, \v{y}) \hat{b}^l_{d}(\v{v}_1)a^{l\dagger}_d(\eta, \v{v}_1)|\gamma^*\rangle\\
%=&\int d^2{x} U^{aa'}_{\v{\bar{u}}_1}U^{bb'}_{\v{\bar{u}}_2}T^{b'}_{a'd}  f^j(\v{\bar{u}}_2-\v{\bar{u}}_1)f^i(\v{\bar{u}}_1-\v{x})\hat\rho_d(\v{x})|\gamma^*\rangle\\
=&\int d^2 \v x U^{aa'}_{\v{\bar{u}}_1}U^{bb'}_{\v{\bar{u}}_2}T^{a'}_{db'}  f^j(\v{\bar{u}}_2-\v{\bar{u}}_1)f^i(\v{\bar{u}}_1-\v{x})\hat\rho_d(\v{x})|\gamma^*\rangle\,.
\end{split}
\end{equation}
Combining these two contributions we complete   ${\cal A}_1$
\begin{equation}
\begin{split}
{\cal A}_1(\bar{\v u}_1,\bar{\v u}_2)=&2\int d^2 \v x f^j(\v{\bar{u}}_2-\v{\bar{u}}_1)f^i(\v{\bar{u}}_1-\v{x}) (U_{\v{\bar{u}}_2}-U_{\v{\bar{u}}_1})^{bb'}[U^{\dagger}_{\v{\bar{u}}_1}T^{a}U_{\v{\bar{u}}_1}]_{b'd}  \hat\rho_d(\v{x})
\end{split}
\end{equation}
where to obtain the final expression  we used the identity
\begin{align}
U^{aa'}_{\v{\bar{u}}_1}T^{a'}_{db'}= [U^{\dagger}_{\v{\bar{u}}_1}T^{a}U_{\v{\bar{u}}_1}]_{db'} \,.
\end{align}

\subsubsection{Amplitude ${\cal A}_2$}
Following similar calculations, we obtain:
\begin{equation}
\begin{split}
&{\cal A}_2(\bar{\v u}_1,\bar{\v u}_2)\\
&=2\int d^2\bm{x}d^2\bm{y}f^i(\bar{\v u}_1-\bm{x})f^j(\bar{\v u}_2-\bm{y})\left[U^{ae}_{\bm{x}}U^{bc}_{\bar{\v u}_2}\hat\rho_e(\bm{x})\hat\rho_{c}(\bm{y})+U^{bc}_{\bm{y}}U^{ae}_{\bar{\v u}_1}\hat\rho_{c}(\bm{y})\hat\rho_{e}(\bm{x})\right.\\
&\left.-U^{ae}_{\bm{x}}U^{bc}_{\bm{y}}\hat\rho_e(\bm{x})\hat\rho_c(\bm{y})-U^{ae}_{\bar{\v u}_1}U^{bc}_{\bar{\v u}_2}\hat{\rho}_{c}(\bm{y})\hat{\rho}_{e}(\bm{x})\right]
\\&= 2\int d^2\bm{x}d^2\bm{y}f^i(\bar{\v u}_1-\bm{x})f^j(\bar{\v u}_2-\bm{y})
\\ &\times \left[U^{ae}_{\bm{x}} \left( U^{bc}_{\bar{\v u}_2}
- U^{bc}_{\bm{y}}
\right)
\hat\rho_e(\bm{x})\hat\rho_{c}(\bm{y})+
U^{ae}_{\bar{\v u}_1}
\left(U^{bc}_{\bm{y}}-U^{bc}_{\bar{\v u}_2}\right)\hat\rho_{c}(\bm{y})\hat\rho_{e}(\bm{x})\right]\,.
\end{split}
\end{equation}
If we were to forget about the quantum nature of the color charge density operators (neglect their commutation relations) the later expression would be identical to the frequently used expression for two gluon production in the dilute-dense limit, see e.g. Appendix A of Ref.~\cite{Kovner:2012jm}.

\subsection{Matrix elements $\Sigma$}
We now compute the matrix elements.
The matrix element proportional to the second power of the projectile charge density is: 
\begin{equation}
\begin{split}
&\Sigma_2(\v u_1, \v u_2, \bar{\v u}_1,\bar{\v u}_2 )= {\cal A}_{1'}^*(\v u_1,\v u_2){\cal A}_1(\bar{\v u}_1,\bar{\v u}_2) \\
&=4\int d^2\bm{x}\int d^2\bar{\bm{x}}f^i(\bar{\v u}_1-\bm{x})f^i(\v u_1-\bar{\bm{x}})f^j(\bar{\v u}_2-\bar{\v u}_1)f^j(\v u_2-\v u_1)\left<\hat \rho_{d'}(\bar{\bm{x}})\hat \rho_{d}(\bm{x})\right>_{P}\\
&\left<\left[ [U^{\dagger}_{\v u_1}T^{a}U_{\ vu_1}][U^{\dagger}_{\v u_2}-U^{\dagger}_{\v u_1}] [U_{\bar{\v u}_2}-U_{\bar{\v u}_1}][U^{\dagger}_{\bar{\v u}_1}T^{a}U_{\bar{\v u}_1}]\right]_{d'd}   \right>_{T}\\
&=4\int d^2\bm{x}\int d^2\bar{\bm{x}}f^i(\bar{\v u}_1-\bm{x})f^i(\v u_1-\bar{\bm{x}})f^j(\bar{\v u}_2-\bar{\v u}_1)f^j(\v u_2-\v u_1)\left<\hat \rho_{d'}(\bar{\bm{x}})\hat \rho_{d}(\bm{x})\right>_{P}\\
&\left<\Tr(U_{\v u_1}T^{e}[U^{\dagger}_{\v u_2}-U^{\dagger}_{\v u_1}][U_{\bar{\v u}_2}-U_{\bar{\v u}_1}]T^eU^{\dagger}_{\bar{\v u}_1})\right>_T
\end{split}
\end{equation}

The interference terms involving the cubic power of the projectile charge density: 
\begin{equation}
\begin{split}
&\Sigma_3(\v u_1, \v u_2, \bar{\v u}_1,\bar{\v u}_2 )= {\cal A}_{2'}^*(\v u_1,\v u_2){\cal A}_1(\bar{\v u}_1,\bar{\v u}_2)+{\cal A}_{1'}^*(\v u_1,\v u_2){\cal A}_{2}(\bar{\v u}_1,\bar{\v u}_2) \\
&=4 \int d^2\bar{\bm{x}}d^2\bar{\bm{y}} d^2\bm{x}f^j(\bar{\v u}_2-\bar{\v u}_1)f^i(\bar{\v u}_1-\bm{x})f^i(\v u_1-\bar{\bm{x}})f^j(\v u_2-\bar{\bm{y}})\\
& \times  \left(\left<\hat{\rho}_{c'}(\bar{\bm{y}})\hat{\rho}_{e'}(\bar{\bm{x}}) \hat{\rho}_{d}(\bm{x})\right>_{P} \right.\\
& \times \left<   U^{\dagger e'a}(\bar{\bm{x}}) \left[ [U^{\dagger}_{\v u_2}-U^{\dagger}_{\bar{\bm{y}}}] [U_{\bar{\v u}_2}-U_{\bar{\v u}_1}][U^{\dagger}_{\bar{\v u}_1}T^{a}U_{\bar{\v u}_1}]\right]_{c'd}                  \right>_T\\
&\left.  +  \left<\hat{\rho}_{e'}(\bar{\bm{x}})\hat{\rho}_{c'}(\bar{\bm{y}}) \hat{\rho}_{d}(\bm{x})\right>_{P}         \right.\\
& \left.   \times \left<   U^{\dagger e'a}(\v u_1) \left[ [U^{\dagger}_{\bar{\bm{y}}}-U^{\dagger}_{\v u_2}] [U_{\bar{\v u}_2}-U_{\bar{\v u}_1}][U^{\dagger}_{\bar{\v u}_1}T^{a}U_{\bar{\v u}_1}]\right]_{c'd}                  \right>_T     \right)                           \\
&+4 \int d^2\bar{\bm{x}}d^2\bm{y} d^2\bm{x}f^j(\v u_2-\v u_1)f^i(\v u_1-\bar{\bm{x}})f^i(\bar{\v u}_1-\bm{x})f^j(\bar{\v u}_2-\bm{y})\\
& \times  \left(\left< \hat{\rho}_{d'}(\bar{\bm{x}})\hat{\rho}_{e}(\bm{x})\hat{\rho}_{c}(\bm{y})\right>_{P} \right.\\
& \times \left<    \left[ [U^{\dagger}_{\v u_1}T^{a}U_{\v u_1}] [U^{\dagger}_{\v u_2}-U^{\dagger}_{\v u_1}][U_{\bar{\v u}_2}-U_{\bm{y}}]\right]_{d'c}    U^{ ae}(\bm{x})              \right>_T\\
&\left.  + \left< \hat{\rho}_{d'}(\bar{\bm{x}})\hat{\rho}_{c}(\bm{y})\hat{\rho}_{e}(\bm{x})\right>_{P}        \right.\\
& \left.   \times \left<    \left[ [U^{\dagger}_{\v u_1}T^{a}U_{\v u_1}] [U^{\dagger}_{\v u_2}-U^{\dagger}_{\v u_1}][U_{\bm{y}}-U_{\bar{\v u}_2}]\right]_{d'c}    U^{ ae}(\bar{\v u}_1)              \right>_T   \right)                          
\end{split}
\end{equation}
%These two terms are merely complex conjugates of each other. 

Finally the term with the highest power of the projectile density is 
\begin{equation}
\begin{split}
&\quad\Sigma_4(\v u_1, \v u_2, \bar{\v u}_1,\bar{\v u}_2 )\\&= {\cal A}_{2'}^*(\v u_1,\v u_2){\cal A}_2(\bar{\v u}_1,\bar{\v u}_2) \\
&=4\int d^2\bm{x}d^2\bm{y}\int d^2\bar{\bm{x}}d^2\bar{\bm{y}}f^i(\bar{\v u}_1-\bm{x})f^j(\bar{\v u}_2-\bm{y})f^i(\v u_1-\bar{\bm{x}})f^j(\v u_2-\bar{\bm{y}})\\
& \times  \left(\left<\hat{\rho}_{c'}(\bar{\bm{y}})\hat{\rho}_{e'}(\bar{\bm{x}}) \hat{\rho}_{e}(\bm{x})\hat{\rho}_{c}(\bm{y})\right>_{P} \left<   [U^{\dagger}_{\bar{\bm{x}}}U_{\bm{x}}]_{e'e} \left[ [U^{\dagger}_{\v u_2}-U^{\dagger}_{\bar{\bm{y}}}][U_{\bar{\v u}_2}-U_{\bm{y}}]\right]_{c'c}                  \right>_T \right.
\\
&\left.+\left<\hat{\rho}_{e'}(\bar{\bm{x}})\hat{\rho}_{c'}(\bar{\bm{y}}) \hat{\rho}_{e}(\bm{x})\hat{\rho}_{c}(\bm{y})\right>_{P} \left<   [U^{\dagger}_{\v u_1}U_{\bm{x}}]_{e'e} \left[ [U^{\dagger}_{\bar{\bm{y}}}-U^{\dagger}_{\v u_2}][U_{\bar{\v u}_2}-U_{\bm{y}}]\right]_{c'c}                  \right>_T \right.
\\
&\left.+\left<\hat{\rho}_{e'}(\bar{\bm{x}})\hat{\rho}_{c'}(\bar{\bm{y}}) \hat{\rho}_{c}(\bm{y})\hat{\rho}_{e}(\bm{x})\right>_{P} \left<   [U^{\dagger}_{\v u_1}U_{\bar{\v u}_1}]_{e'e} \left[ [U^{\dagger}_{\bar{\bm{y}}}-U^{\dagger}_{\v u_2}][U_{\bm{y}}-U_{\bar{\v u}_2}]\right]_{c'c}                  \right>_T \right.\\
&\left.+\left<\hat{\rho}_{c'}(\bar{\bm{y}})\hat{\rho}_{e'}(\bar{\bm{x}}) \hat{\rho}_{c}(\bm{y})\hat{\rho}_{e}(\bm{x})\right>_{P} \left<   [U^{\dagger}_{\bar{\bm{x}}}U_{\bar{\v u}_1}]_{e'e} \left[ [U^{\dagger}_{\v u_2}-U^{\dagger}_{\bar{\bm{y}}}][U_{\bm{y}}-U_{\bar{\v u}_2}]\right]_{c'c}                  \right>_T \right)
\end{split}
\end{equation}

\section{Inclusive two gluon production: the  dipole model}\label{sec4}
\label{Sect:2glDPM}
%First, explicit transformation to momentum space can easily account for translational  invariance and thus reduce the number of ensuing integrals. Second, multi-dimensional integrals with multiple Fourier phases are notoriously difficult for the Monte-Carlo integration.  The transformation into the momentum space  allows to take care of this difficulty analytically. 

The two particle inclusive cross section in terms of the matrix element $\Sigma = \Sigma_2 +  \Sigma_3 +  \Sigma_4$ computed previously, reads 
\begin{equation}
    \begin{split}
        \frac{dN}{d\eta d \v{q}_1^2 d\xi d \v{q}_2^2}&=\frac{1}{4} \sum_{s_1,s_2,\lambda}\int_{0}^{1}\frac{dz}{2\pi z(1-z)} \int_{\v z_1,\v z_2} \Psi^{T*}_{\lambda}(z,\v z_1-\v z_2,s_1)\Psi^{T}_{\lambda}(z,\v z_1-\v z_2,s_1)\\ & \times \frac{1}{(2\pi)^4} \int_{\v u_1, \bar{\v u}_1,\v u_2, \bar{\v u}_2} e^{-i \v{q}_1(\v u_1-\bar{\v u}_1)}e^{-i \v{q}_2(\v u_2-\bar{\v u}_2)}\Sigma(\v u_1, \v u_2, \bar{\v u}_1,\bar{\v u}_2 )\,.
    \end{split}
\end{equation}
%where the photon splitting function  in the coordinate space is given by 
%\begin{align}
%\Psi^{T}_{\lambda}(z, \v r ,s_1)&=-i\frac{2ee_f}{2\pi}\delta_{s_1,-s_2}(2z-1+2\lambda s_1)\sqrt{z(1-z)}\frac{\v r \cdot \v{\epsilon_{\lambda} }}{|\v r|}\varepsilon_fK_1(\varepsilon_f|\v r|)\,.
%\end{align}
%As we alluded to before, here we consider only the dominant, transverse polarization of the near-real photon. 

Summing over the quark and (transverse) photon polarizations  we obtain 
\begin{align}
    \sum_{s_1,s_2,\lambda}|\Psi^{T}_{\lambda}(z,\v z_1-\v z_2,s_1)|^2
    =&\frac{32e^2e_f^2}{(2\pi)^2}(z^2+ {\bar z}^2)z \bar z \varepsilon^2_fK_1(\varepsilon_f|\v r|)^2 
\end{align}
where we introduced $\bar z=1-z$ to simplify notation. 

%In principle $\varepsilon_f$ is a function of $z$; however for small $\varepsilon_f \propto Q$, the combination $\varepsilon_f^2 K_1^2(\varepsilon_f  r)$ is approximately $\varepsilon_f$ and therefore $z$ independent. We thus set $z$ to $1/2$ in  $\varepsilon_f^2 K_1^2(\varepsilon_f  r)$ and 
We now integrate over $z$ explicitly: 
\begin{align}
    \int^1_0 dz \, (z^2+{\bar z}^2)=\frac{2}{3}\,.
\end{align}
Hence the cross section reads 
\begin{equation}
    \begin{split}
        \frac{dN}{d\eta d \v{q}_1^2 d\xi d \v{q}_2^2}&=\frac{16e^2e_f^2}{3 (2\pi)^3}\int_{\bm{z}_1,\bm{z}_2}\varepsilon^2_fK_1(\varepsilon_f|\bm{z}_1-\bm{z}_2|)^2 \\
        &\times \frac{1}{(2\pi)^4} \int_{u_1, \bar{u}_1,u_2, \bar{u}_2} e^{-i \v{q}_1(\v u_1-\bar{\v u}_1)}e^{-i \v{q}_2(\v u_2-\bar{\v u}_2)}\Sigma(\v u_1, \v u_2, \bar{\v u}_1,\bar{\v u}_2 ) 
    \end{split}
\end{equation}
where as discussed earlier we set $\varepsilon_f = Q/2$. 

Numerical calculations are easier performed in  momentum space, therefore in the following we present all expressions in terms of momentum integrals.

It is convenient to split the production cross section into three terms corresponding to $\Sigma_2$, $\Sigma_3$ and $\Sigma_4$.
Using the correlators of the color charge density in the dipole, calculated in Section 2, we obtain for the contribution of $\Sigma_2$:
\begin{equation}
\label{Eq:N2}
    \begin{split}
             \frac{dN^{(2)}}{d\eta d\v{q}_1^2 d\xi d\v{q}_2^2}&=\frac{16 e^2e_f^2}{3 \pi^4} \alpha_s^2\int_{P, \v{k} ,\v{\bar{k}} } \frac{\mathcal{I}(\varepsilon_f,P)}{P^2}\, \Gamma(\v{q}_2,\v k,\v{\bar{k}})
             \\
         &\Tr(U_{\v{q}_1+\v{k}-\v{P}}T^{e}U^{\dagger}_{\v{k}-\v{q}_2}U_{\v{\bar{k}}-\v{q}_2}T^eU^{\dagger}_{\v{q}_1+\v{\bar{k}} -\v{P}})\,.
    \end{split}
\end{equation}
%To obtain this expression we used simplified the coordinate space expression first  
%\begin{equation}
 %   \begin{split}
  %      &\Tr\left[ U^{\dagger}_{\v u_1}T^{a}U_{\v u_1}[U^{\dagger}_{\v u_2}-U^{\dagger}_{\v u_1}][U_{\bar{\v u}_2}-U_{\bar{\v u}_1}]U^{\dagger}_{\bar{\v u}_1}T^{a}U_{\bar{\v u}_1}\right]\\
   %    =&\Tr(U_{\v u_1}T^{e}[U^{\dagger}_{\v u_2}-U^{\dagger}_{\v u_1}][U_{\bar{\v u}_2}-U_{\bar{\v u}_1}]T^eU^{\dagger}_{\bar{\v u}_1})
    %\end{split}
%\end{equation}
%which can be proven by using  the identity $U^{aa'}_{\v{\bar{u}}_1}T^{a'}_{db'}= [U^{\dagger}_{\v{\bar{u}}_1}T^{a}U_{\v{\bar{u}}_1}]_{db'}$\,.
In Eq.~\eqref{Eq:N2} we introduced 
\begin{align}
    \mathcal{I}_{\varepsilon_f}(|\v P|) &= 
    \int_0^{\infty} dr r \epsilon_f^2 K^2_1(\epsilon_f |\v r|)(1-J_0(|\v r||\v P|))\notag \\ &= 
    -\frac{1}{2}+\frac{(\v P^2+2\epsilon_f^2)\ln(\frac{|\v P|}{2\epsilon_f}+\sqrt{1+\frac{\v P^2}{4 \epsilon_f^2}}) }{|\v P|\sqrt{\v P^2+4\epsilon_f^2}} ,
    \\ 
    \Gamma (\v{q}_2,\v{k},\v{\bar{k}}) &= \left(\frac{\v q^i_2}{\v{q}_2^2}-\frac{ \v k^i}{ \v k^2}\right) \left(\frac{\v q^i_2}{\v{q}_2^2}-\frac{\v{\bar{k}}^i}{\v{\bar{k}}^2}\right)\,.
\end{align}

%One can check that $ \mathcal{I}(\varepsilon_f,P)$ has no infrared divergence,
%\begin{equation}
%    \begin{split}
%        \lim_{P\rightarrow 0} \mathcal{I}(\varepsilon_f,P)=\lim_{P\rightarrow 0}-\frac{1}{2}+\frac{(P^2+2\epsilon_f^2)\frac{P}{2\epsilon_f} }{|P|\sqrt{P^2+4\epsilon_f^2}} =0
%    \end{split}
%\end{equation}
Note that the ratio $\mathcal{I}(\varepsilon_f,P) / P^2$ as a function of $P^2$ approaches a finite constant in the IR($P/\varepsilon_f\to0$). 
%To check the final expression, it is useful to consider the limit of trivial scattering when all Wilson lines in the coordinate space are replaced by $1$ and in momentum space by the Dirac delta function of the argument. In this limit the production cross section must vanish. One can see that Eq.~\eqref{Eq:N2} trivially satisfies this requirement, as $\Gamma(\v{q}_2,\v k,\v{\bar{k}})$ goes to zero when either (or both) $\v k$ and $\v{\bar{k}}$ are equal to $\v{q}_2$.  

The contributions involving $\Sigma_3$ cannot be simplified into one homogeneous expression.
Instead we have
\begin{equation}
\frac{dN^{(3)}}{d\eta d\v{q}_1^2 d\xi d\v{q}_2^2}=\frac{dN_1^{(3)}}{d\eta d\v{q}_1^2 d\xi d\v{q}_2^2}+\frac{dN_2^{(3)}}{d\eta d\v{q}_1^2 d\xi d\v{q}_2^2}+\frac{dN_3^{(3)}}{d\eta d\v{q}_1^2 d\xi d\v{q}_2^2}+\frac{dN_4^{(3)}}{d\eta d\v{q}_1^2 d\xi d\v{q}_2^2}
\end{equation}
We list all four terms separately. 
\begin{equation}
    \begin{split}
       \frac{dN_1^{(3)}}{d\eta d\v{q}_1^2 d\xi d\v{q}_2^2}
=& %\frac{32e^2e_f^2g^4}{3}\frac{T_{c'e'}^d}{(2\pi)^6} 
- \frac{8 e^2e_f^2}{3 \pi^4} \alpha_s^2
\int_{\v{k}_1,\v{k}_2,\v{k}_3} \left\langle \Tr \left[T^d U^{\dagger}_{\v{k}_1}U_{\v{k}_2}T^dU^{\dagger}_{\v{k}_3}U_{\v{k}_1-\v{k}_2+\v{k}_3}\right]\right\rangle_T \\
&\times\frac{\v{q}_1 \cdot (\v{q}_1+\v{q}_2+\v{k}_2-\v{k}_3)}{\v{q}_1^2(\v{q}_1+\v{q}_2+\v{k}_2-\v{k}_3)^2}\Gamma(\v{q}_2,\v{q}_2+\v{k}_1,\v{q}_2+\v{k}_2)\\
&\times\Big[\mathcal{I}_{\varepsilon_f}( |\v{q}_1+\v{q}_2+\v{k}_2-\v{k}_3|)+\mathcal{I}_{\varepsilon_f}( |\v{q}_2+\v{k}_1|)-\mathcal{I}_{\varepsilon_f}( |\v{q}_1-\v{k}_1+\v{k}_2-\v{k}_3|)\Big]\,.
    \end{split}
\end{equation}
Note that this expression does not have poles at $\v{q}_2+\v{k}=0$ and $\v{q}_1+\v{q}_2+\v{k}_2-\v{k}_3=0$  as the combination of $\mathcal{I}$ functions vanishes at these points. 

Similarly for the remaining three contributions we have 
\begin{equation}
    \begin{split}
       \frac{dN_2^{(3)}}{d\eta d\v{q}_1^2 d\xi d\v{q}_2^2}&=-
       \frac{8 e^2e_f^2}{3 \pi^4} \alpha_s^2
       \int_{\v{k}_1,\v{k}_2,\v{k}_3}\left\langle \Tr \left[T^d U^{\dagger}_{\v{k}_1}U_{\v{k}_2}T^dU^{\dagger}_{\v{k}_3}U_{\v{k}_1-\v{k}_2+\v{k}_3} \right]\right\rangle_T \\
&\times\frac{(\v{q}_1-\v{k}_1+\v{k}_2-\v{k}_3) \cdot
(\v{q}_1+\v{q}_2+\v{k}_2-\v{k}_3)}{(\v{q}_1-\v{k}_1+\v{k}_2-\v{k}_3)^2(\v{q}_1+\v{q}_2+\v{k}_2-\v{k}_3)^2} \Gamma(\v{q}_2,\v{q}_2+\v{k}_1,\v{q}_2+\v{k}_2) \\
&\times \Big[\mathcal{I}_{\varepsilon_f}( |\v{q}_1-\v{k}_1+\v{k}_2-\v{k}_3|)+\mathcal{I}_{\varepsilon_f}( |\v{q}_1+\v{q}_2+\v{k}_2-\v{k}_3|)-\mathcal{I}_{\varepsilon_f}( |\v{q}_2+\v{k}_1|)\Big]\,,
    \end{split}
\end{equation}
\begin{equation}
    \begin{split}
       \frac{dN_3^{(3)}}{d\eta d\v{q}_1^2 d\xi d\v{q}_2^2}&=
       - \frac{8 e^2e_f^2}{3 \pi^4} \alpha_s^2
       \int_{\v{k}_1,\v{k}_2,\v{k}_3}
       \left \langle \Tr \left [T^{d} U^{\dagger}_{\v{k}_1}U_{\v{k}_2}T^{d}U^{\dagger}_{\v{k}_3}U_{\v{k}_1-\v{k}_2+\v{k}_3} \right] \right\rangle_T \\
&\times\frac{\v{q}_1\cdot (\v{q}_1+\v{q}_2-\v{k}_2+\v{k}_3)}{\v{q}_1^2(\v{q}_1+\v{q}_2-\v{k}_2+\v{k}_3)^2}\Gamma(\v{q}_2,\v{q}_2+\v{k}_3,\v{q}_2+\v{k}_1-\v{k}_2+\v{k}_3)\\
&\times \Big[\mathcal{I}_{\varepsilon_f}( |\v{q}_2+\v{k}_1-\v{k}_2+\v{k}_3|)+\mathcal{I}_{\varepsilon_f}( |\v{q}_1+\v{q}_2-\v{k}_2+\v{k}_3|)-\mathcal{I}_{\varepsilon_f}( |\v{q}_1-\v{k}_1|)\Big]\,,
    \end{split}
\end{equation}
and, finally, 
\begin{equation}
    \begin{split}
       \frac{dN_4^{(3)}}{d\eta d\v{q}_1^2 d\xi d\v{q}_2^2}&=
       - \frac{8 e^2e_f^2}{3 \pi^4} \alpha_s^2
       \int_{\v{k}_1,\v{k}_2,\v{k}_3}\left\langle \Tr \left[T^d U^{\dagger}_{\v{k}_1}U_{\v{k}_2}T^{d}U^{\dagger}_{\v{k}_3}U_{\v{k}_1-\v{k}_2+\v{k}_3}\right]\right\rangle_T \\
&\times\frac{(\v{q}_1-\v{k}_1)\cdot (\v{q}_1+\v{q}_2-\v{k}_2+\v{k}_3)}{(\v{q}_1-\v{k}_1)^2(\v{q}_1+\v{q}_2-\v{k}_2+\v{k}_3)^2}\Gamma(\v{q}_2,\v{q}_2+\v{k}_3,\v{q}_2+\v{k}_1-\v{k}_2+\v{k}_3)\\
&\times \Big[\mathcal{I}_{\varepsilon_f}( |\v{q}_1-\v{k}_1|)+\mathcal{I}_{\varepsilon_f}( |\v{q}_1+\v{q}_2-\v{k}_2+\v{k}_3|)-\mathcal{I}_{\varepsilon_f}( |\v{q}_2+\v{k}_1-\v{k}_2+\v{k}_3|)\Big]\,.
    \end{split}
\end{equation}

Switching to the contribution from $\Sigma_4$ involving four $\rho(x)$ correlators, we again obtain four terms that cannot be obviously combined:
\begin{equation}
\frac{dN^{(4)}}{d\eta d\v{q}_1^2 d\xi d\v{q}_2^2}=\frac{dN_1^{(4)}}{d\eta d\v{q}_1^2 d\xi d\v{q}_2^2}+\frac{dN_2^{(4)}}{d\eta d\v{q}_1^2 d\xi d\v{q}_2^2}+\frac{dN_3^{(4)}}{d\eta d\v{q}_1^2 d\xi d\v{q}_2^2}+\frac{dN_4^{(4)}}{d\eta d\v{q}_1^2 d\xi d\v{q}_2^2}
\end{equation}
The explicit expressions are:
\begin{equation}
    \begin{split}
       &\frac{dN_1^{(4)}}{d\eta d\v{q}_1^2 d\xi d\v{q}_2^2}\\&= \frac{32 e^2e_f^2}{3 \pi^4}  \alpha_s^2 \int_{\v{k}_1,\v{k}_2,\v{k}_3}\frac{1}{\v{q}_1^2}\Gamma(\v{q}_2,\v{q}_2+\v{k}_3,\v{q}_2+\v{k}_1-\v{k}_2+\v{k}_3)
       \\
       &\times[U^{\dagger}_{\v{k}_1}U_{\v{k}_2}]_{e'e}  \left[U^{\dagger}_{\v{k}_3}U_{\v{k}_1+\v{k}_3-\v{k}_2}\right]_{c'c}
       \\
       &\times\Big[\Tr(t^{c'}t^{e'}t^{e}t^{c})\Big(\mathcal{I}_{\varepsilon_f}(|\v{q}_2+\v{k}_3|)+\mathcal{I}_{\varepsilon_f}(|\v{q}_1+\v{k}_1|)+\mathcal{I}_{\varepsilon_f}(|\v{q}_2+\v{k}_1-\v{k}_2+\v{k}_3|)
        \\
        +&\mathcal{I}_{\varepsilon_f}(|\v{q}_1+\v{k}_2|)-\mathcal{I}_{\varepsilon_f}(|\v{k}_1-\v{k}_2|)-\mathcal{I}_{\varepsilon_f}(|\v{q}_1-\v{q}_2+\v{k}_2-\v{k}_3|)-\mathcal{I}_{\varepsilon_f}(|\v{q}_1+\v{q}_2+\v{k}_1+\v{k}_3|)\Big)
        \\
        +&\frac{T_{ec}^aT_{c'e'}^a}{8}
        \Big(\mathcal{I}_{\varepsilon_f}(|\v{q}_2+\v{k}_3|)-\mathcal{I}_{\varepsilon_f}(|\v{q}_1+\v{k}_1|)+\mathcal{I}_{\varepsilon_f}(|\v{q}_2+\v{k}_1-\v{k}_2+\v{k}_3|)
        \\
        -&\mathcal{I}_{\varepsilon_f}(|\v{q}_1+\v{k}_2|)-\mathcal{I}_{\varepsilon_f}(|\v{k}_1-\v{k}_2|)+\mathcal{I}_{\varepsilon_f}(|\v{q}_1-\v{q}_2+\v{k}_2-\v{k}_3|)+\mathcal{I}_{\varepsilon_f}(|\v{q}_1+\v{q}_2+\v{k}_1+\v{k}_3|)\Big)
        \Big]
    \end{split}
\end{equation}
One in principle can substitute an explicit expression for the trace of four Gell-Mann matrices, in practice however, this does not help to simplify the expression. 

\begin{equation}
    \begin{split}
       &\frac{dN_2^{(4)}}{d\eta d\v{q}_1^2 d\xi d\v{q}_2^2}\\=&-\frac{32 e^2e_f^2}{3 \pi^4}  \alpha_s^2 \int_{\v{k}_1,\v{k}_2,\v{k}_3}\frac{\v{q}_1 \cdot(\v{q}_1+\v{k}_1)}{\v{q}_1^2(\v{q}_1+\v{k}_1)^2}\Gamma(\v{q}_2,\v{q}_2+\v{k}_3,\v{q}_2+\v{k}_1+\v{k}_3-\v{k}_2)
       \\
       &\times[U^{\dagger}_{\v{k}_1}U_{\v{k}_2}]_{e'e}  \left[U^{\dagger}_{\v{k}_3}U_{\v{k}_1+\v{k}_3-\v{k}_2}\right]_{c'c}
       \\
       &\times\Big[\Tr(t^{e'}t^{c'}t^{e}t^{c})\Big(\mathcal{I}_{\varepsilon_f}(|\v{q}_2+\v{k}_3|)+\mathcal{I}_{\varepsilon_f}(|\v{q}_1+\v{k}_1|)+\mathcal{I}_{\varepsilon_f}(|\v{q}_2+\v{k}_1-\v{k}_2+\v{k}_3|)
        \\
        +&\mathcal{I}_{\varepsilon_f}(|\v{q}_1+\v{k}_2|)-\mathcal{I}_{\varepsilon_f}(|\v{k}_1-\v{k}_2|)-\mathcal{I}_{\varepsilon_f}(|\v{q}_1-\v{q}_2+\v{k}_2-\v{k}_3|)-\mathcal{I}_{\varepsilon_f}(|\v{q}_1+\v{q}_2 + \v{k}_1+\v{k}_3|)\Big)
        \\
        +&\frac{T_{ec}^aT_{e'c'}^a}{8}
        \Big(\mathcal{I}_{\varepsilon_f}(|\v{q}_2+\v{k}_3|)+\mathcal{I}_{\varepsilon_f}(|\v{q}_1+\v{k}_1|)+\mathcal{I}_{\varepsilon_f}(|\v{q}_2+\v{k}_1-\v{k}_2+\v{k}_3|)
        \\
        -&\mathcal{I}_{\varepsilon_f}(|\v{q}_1+\v{k}_2|)+\mathcal{I}_{\varepsilon_f}(|\v{k}_1-\v{k}_2|)-\mathcal{I}_{\varepsilon_f}(|\v{q}_1-\v{q}_2+\v{k}_2-\v{k}_3|)-\mathcal{I}_{\varepsilon_f}(|\v{q}_1+\v{q}_2 + \v{k}_1+\v{k}_3|)\Big)\Big]
    \end{split}
\end{equation}
\begin{equation}
    \begin{split}
       &\frac{dN_3^{(4)}}{d\eta d\v{q}_1^2 d\xi d\v{q}_2^2}\\=& 
       \frac{32 e^2e_f^2}{3 \pi^4}  \alpha_s^2
       \int_{\v{k}_1,\v{k}_2,\v{k}_3}\frac{(\v{q}_1+\v{k}_2) \cdot(\v{q}_1+\v{k}_1)}{(\v{q}_1+\v{k}_2)^2(\v{q}_1+\v{k}_1)^2}\Gamma(\v{q}_2,\v{q}_2+\v{k}_3,\v{q}_2+\v{k}_1+\v{k}_3-\v{k}_2)
       \\
       &\times[U^{\dagger}_{\v{k}_1}U_{\v{k}_2}]_{e'e}  \left[U^{\dagger}_{\v{k}_3}U_{\v{k}_1+\v{k}_3-\v{k}_2}\right]_{c'c}
       \\
       &\times\Big[\Tr(t^{e'}t^{c'}t^{c}t^{e})\Big(\mathcal{I}_{\varepsilon_f}(|\v{q}_2+\v{k}_3|)+\mathcal{I}_{\varepsilon_f}(|\v{q}_1+\v{k}_1|)+\mathcal{I}_{\varepsilon_f}(|\v{q}_2+\v{k}_1-\v{k}_2+\v{k}_3|)
        \\
        +&\mathcal{I}_{\varepsilon_f}(|\v{q}_1+\v{k}_2|)-\mathcal{I}_{\varepsilon_f}(|\v{k}_1-\v{k}_2|)-\mathcal{I}_{\varepsilon_f}(|\v{q}_1-\v{q}_2+\v{k}_2-\v{k}_3|)-\mathcal{I}_{\varepsilon_f}(|\v{q}_1+\v{q}_2 + \v{k}_1+\v{k}_3|)\Big)
        \\
        &\frac{T_{ce}^aT_{e'c'}^a}{8}\Big(-\mathcal{I}_{\varepsilon_f}(|\v{q}_2+\v{k}_3|)+\mathcal{I}_{\varepsilon_f}(|\v{q}_1+\v{k}_1|)+\mathcal{I}_{\varepsilon_f}(|\v{q}_2+\v{k}_1-\v{k}_2+\v{k}_3|)
        \\
        &+\mathcal{I}_{\varepsilon_f}(|\v{q}_1+\v{k}_2|)-\mathcal{I}_{\varepsilon_f}(|\v{k}_1-\v{k}_2|)-\mathcal{I}_{\varepsilon_f}(|\v{q}_1-\v{q}_2+\v{k}_2-\v{k}_3|)+\mathcal{I}_{\varepsilon_f}(|\v{q}_1+\v{q}_2 + \v{k}_1+\v{k}_3|)\Big)\Big]
    \end{split}
\end{equation}
\begin{equation}
    \begin{split}
       &\frac{dN_4^{(4)}}{d\eta d\v{q}_1^2 d\xi d\v{q}_2^2}\\=&-\frac{32 e^2e_f^2}{3 \pi^4}  \alpha_s^2 \int_{\v{k}_1,\v{k}_2,\v{k}_3}\frac{(\v{q}_1+\v{k}_2) \cdot \v{q}_1}{(\v{q}_1+\v{k}_2)^2\v{q}_1^2}\Gamma(\v{q}_2,\v{q}_2+\v{k}_3,\v{q}_2+\v{k}_1+\v{k}_3-\v{k}_2)
       \\
       &\times[U^{\dagger}_{\v{k}_1}U_{\v{k}_2}]_{e'e}  \left[U^{\dagger}_{\v{k}_3}U_{\v{k}_1+\v{k}_3-\v{k}_2}\right]_{c'c}
       \\
       &\times\Big[\Tr(t^{c'}t^{e'}t^{c}t^{e})\Big(\mathcal{I}_{\varepsilon_f}(|\v{q}_2+\v{k}_3|)+\mathcal{I}_{\varepsilon_f}(|\v{q}_1+\v{k}_1|)+\mathcal{I}_{\varepsilon_f}(|\v{q}_2+\v{k}_1-\v{k}_2+\v{k}_3|)
        \\
        +&\mathcal{I}_{\varepsilon_f}(|\v{q}_1+\v{k}_2|)-\mathcal{I}_{\varepsilon_f}(|\v{k}_1-\v{k}_2|)-\mathcal{I}_{\varepsilon_f}(|\v{q}_1-\v{q}_2+\v{k}_2-\v{k}_3|)-\mathcal{I}_{\varepsilon_f}(|\v{q}_1+\v{q}_2 + \v{k}_1+\v{k}_3|)\Big)
        \\
        +&\frac{T_{ce}^aT_{c'e'}^a}{8}\Big(\mathcal{I}_{\varepsilon_f}(|\v{q}_2+\v{k}_3|)-\mathcal{I}_{\varepsilon_f}(|\v{q}_1+\v{k}_1|)-\mathcal{I}_{\varepsilon_f}(|\v{q}_2 + \v{k}_1-\v{k}_2 + \v{k}_3|)
        \\
        &+\mathcal{I}_{\varepsilon_f}(|\v{q}_1+\v{k}_2|)+\mathcal{I}_{\varepsilon_f}(|\v{k}_1-\v{k}_2|)-\mathcal{I}_{\varepsilon_f}(|\v{q}_1-\v{q}_2+\v{k}_2-\v{k}_3|)+\mathcal{I}_{\varepsilon_f}(|\v{q}_1+\v{q}_2+\v{k}_1+\v{k}_3|)\Big)\Big]
    \end{split}
\end{equation}
Note that the combinations of the functions $\mathcal{I}$ multiplying the trace of the four Gell-Mann matrices are identical, while those multiplying the product of two adjoint matrices are different in different terms due to distinct patterns of signs.  

It is interesting to separate out the contribution which is analogous to calculations performed in the ``classical'' MV model. As we have discussed above, it corresponds to keeping only those terms in $\Sigma_4$ which are completely symmetric in the color charge density operators. Here this corresponds to symmetrizing the four generators in the trace with respect to the color indices.
For this symmetric part of the trace, see Eq.~\eqref{Eq:TrSym}, and $N_c=3$ we obtain  
\begin{equation}
    \begin{split}
       \frac{dN_{\rm sym}^{(4)}}{d\eta d\v{q}_1^2 d\xi d\v{q}_2^2}
        =&\frac{16 e^2e_f^2}{27 \pi^4}  \alpha_s^2  \int_{\v{k}_1,\v{k}_2,\v{k}_3}\Gamma(\v{q}_1,\v{q}_1+\v{k}_1,\v{q}_1+\v{k}_2)\Gamma(\v{q}_2,\v{q}_2+\v{k}_3,\v{q}_2+\v{k}_1-\v{k}_2+\v{k}_3)
        \\
        &\times\Big[\Big(\mathcal{I}_{\varepsilon_f}(|\v{q}_2+\v{k}_3|)+\mathcal{I}_{\varepsilon_f}(|\v{q}_1+\v{k}_1|)+\mathcal{I}_{\varepsilon_f}(|\v{q}_2+\v{k}_1-\v{k}_2+\v{k}_3|)
        \\
        &\quad \quad +\mathcal{I}_{\varepsilon_f}(|\v{q}_1+\v{k}_2|)-\mathcal{I}_{\varepsilon_f}(|\v{k}_1-\v{k}_2|)
        \\
        &\quad \quad -
        \mathcal{I}_{\varepsilon_f}(|\v{q}_1-\v{q}_2+\v{k}_2-\v{k}_3|)-\mathcal{I}_{\varepsilon_f}(|\v{q}_1+\v{q}_2+\v{k}_1+\v{k}_3|)\Big)
        \Big]\\
        &\Big(  \Tr[U^{\dagger}_{\v{k}_1}U_{\v{k}_2}]\Tr\left[U^{\dagger}_{\v{k}_3}U_{\v{k}_1+\v{k}_3-\v{k}_2}\right] 
        \\
        &\quad  + 
        \Tr\left[U^{\dagger}_{\v{k}_1}U_{\v{k}_2}U^{\dagger}_{\v{k}_3}U_{\v{k}_1+\v{k}_3-\v{k}_2}\right]
        \\
        &\quad +     
         \Tr\left[U^{\dagger}_{\v{k}_1}U_{\v{k}_2}U^{\dagger}_{-\v{k}_1+\v{k}_2-\v{k}_3}U_{-\v{k}_3}\right]          \Big)\,.
    \end{split}
\end{equation}

This expression has a few physically transparent  properties. First, it vanishes if any of the four interactions with the target is set to zero, $U_{\v{k}} \to \delta^{(2)}(\v{k})$, due to  $\Gamma(\v q,\v k,\v q) = \Gamma(\v q,\v q,\v k)=0$. Second, the color neutrality of the dipole suggests that  a zero momentum gluon will not be able to resolve the dipole and thus lead to a trivial contribution. Gluons originating from the dipole have the momenta $\v{q}_1+\v{k}_1$, $\v{q}_1+\v{k}_2$, $\v{q}_2+\v{k}_3$, and $\v{q}_2+\v{k}_1-\v{k}_2+\v{k}_3$ (this can be read off the arguments of $\Gamma$ functions). Setting any of this to zero leads to vanishing contribution in the integral due to the combination of $\mathcal{I}$ functions.   
Finally, one can trivially show that the cross section is symmetric under $\v{q}_2\to -\v{q}_2$.

\subsection{Averaging over the target in FDA}
 We now need to average over the target field ensemble using the  FDA. 
We present here the detailed computation only for the term $\Sigma_2$. We have 
\begin{equation}
    \begin{split}
      &\langle \Tr(U_{\v{q}_1+\v{k}-\v{P}}T^{e}U^{\dagger}_{\v{k}-\v{q}_2}U_{\v{\bar{k}}-\v{q}_2}T^eU^{\dagger}_{\v{q}_1+\v{\bar{k}} -\v{P}})\rangle_T\\
      =&T^e_{bc}T^e_{fg} \langle U^{ab}_{\v{q}_1+\v{k}-\v{P}} U^{\dagger cd}_{\v{k}-\v{q}_2} \rangle_T \langle U^{df}_{\v{\bar{k}}-\v{q}_2} U^{\dagger ga}_{\v{q}_1+\v{\bar{k}} -\v{P}})\rangle_T\\
      +&T^e_{bc}T^e_{fg} \langle U^{ab}_{\v{q}_1+\v{k}-\v{P}}U^{df}_{\v{\bar{k}}-\v{q}_2}  \rangle_T \langle U^{\dagger cd}_{\v{k}-\v{q}_2}  U^{\dagger ga}_{\v{q}_1+\v{\bar{k}} -\v{P}})\rangle_T\\
      +&T^e_{bc}T^e_{fg} \langle U^{ab}_{\v{q}_1+\v{k}-\v{P}} U^{\dagger ga}_{\v{q}_1+\v{\bar{k}} -\v{P}}) \rangle_T \langle U^{\dagger cd}_{\v{k}-\v{q}_2} U^{df}_{\v{\bar{k}}-\v{q}_2} \rangle_T\\
    \end{split}
\end{equation}
The first term vanishes since $T^e_{bc} \delta_{bc}=0$. The second term  
\begin{equation}
    \begin{split}
        &T^e_{bc}T^e_{fg} \langle U^{ab}_{\v{q}_1+\v{k}-\v{P}}U^{df}_{\v{\bar{k}}-\v{q}_2}  \rangle_T \langle U^{\dagger cd}_{\v{k}-\v{q}_2}  U^{\dagger ga}_{\v{q}_1+\v{\bar{k}} -\v{P}})\rangle_T\\
        &=T^e_{bc}T^e_{b c}\frac{(2\pi)^4}{N_c^2-1} \delta^{(2)}(0) \delta^{(2)}(\v{q}_1+\v{k}+\v{\bar{k}}-\v{P}-\v{q}_2)D(\v{q}_2-\v{\bar{k}})D(\v{q}_2- \v k) 
        \\
        & = 
     - (2\pi)^2 N_c S_\perp  \delta^{(2)}(\v{q}_1+\v{k}+\v{\bar{k}}-\v{P}-\v{q}_2)D(\v{q}_2-\v{\bar{k}})D(\v{q}_2- \v k) , 
    \end{split}
\end{equation}
where we took into account that in the momentum space $\delta^{(2)}(0) = S_\perp/(2\pi)^2$. 
The third term  
\begin{equation}
    \begin{split}
        &T^e_{bc}T^e_{fg} \langle U^{ab}_{\v{q}_1+\v{k}-\v{P}} U^{\dagger ga}_{\v{q}_1+\v{\bar{k}} -\v{P}}) \rangle_T \langle U^{\dagger cd}_{\v{k}-\v{q}_2} U^{df}_{\v{\bar{k}}-\v{q}_2} \rangle_T\\
    &= (2\pi)^2 \Tr (T^e T^e) S_\perp 
    \delta^{(2)}(\v{k}-\v{\bar{k}})D(\v{q}_1+\v{k}-\v{P})D(\v{q}_2-\v{k})\\ 
    &= (2\pi)^2 N_c (N_c^2-1) S_\perp 
    \delta^{(2)}(\v{k}-\v{\bar{k}})D(\v{q}_1+\v{k}-\v{P})D(\v{q}_2-\v{k})\,.
    \end{split}
\end{equation}
Hence the cross section for the contribution involving $\Sigma_2$ is 
\begin{equation}
    \begin{split}
        \frac{dN^{(2)}}{d\eta d\v{q}_1^2 d\xi d\v{q}_2^2}=&\frac{64e^2e_f^2 }{3\pi^2} N_c  \alpha_s^2 S_{\perp} \int_{k,\v{\bar{k}} } \frac{\mathcal{I}(\varepsilon,|\v{q}_1+\v{q}_2+\v{k}+\v{\bar{k}}|)}{(\v{q}_1+\v{q}_2+\v{k}+\v{\bar{k}})^2}\\
        &\Big((N_c^2-1)\Gamma(\v{q}_2,\v{q}_2+\v{k},\v{q}_2+\v{k})- \Gamma(\v{q}_2,\v{q}_2+ k,\v{q}_2+ \v{\bar{k}})\Big) D(\v{\bar{k}})D(\v{k})\,.
    \end{split}
\end{equation}
Similarly we obtain 
\begin{equation}
    \begin{split}
         \frac{dN^{(3)}}{d\eta d\v{q}_1^2 d\xi d\v{q}_2^2}
         &=\frac{32 e^2e_f^2}{3 \pi^2} N_c\alpha_s^2 S_{\perp} \int_{k,\v{\bar{k}}}
         \Big[\mathcal{I}_{\varepsilon_f}(|\v{q}_1+\v{q}_2+\v{k}-\v{\bar{k}}|) L(\v{q}_1, \v{\bar{k}}-\v{q}_1)\\
         &+\Big(\mathcal{I}_{\varepsilon_f}(|\v{k}+\v{q}_2|)-\mathcal{I}_{\varepsilon_f}(|\v{\bar{k}}-\v{q}_1|)\Big) L(\v{q}_1, \v{q}_1-\v{\bar{k}})\Big] \cdot \frac{
          (\v{\bar{k}}-\v{k}-\v{q}_1-\v{q}_2)}{(\v{\bar{k}}-\v{k}_1-\v{q}_1-\v{q}_2)^2}
         \\ &\times 
          \Big((N_c^2-1)\Gamma(\v{q}_2,\v{q}_2+\v{k},\v{q}_2+\v{k})-\Gamma(\v{q}_2,\v{q}_2+\v{k},\v{q}_2-\v{\bar{k}})\Big)
         D(\v{k})D(\v{\bar{k}} )
   \end{split}
\end{equation}
For the symmetric part of $\Sigma_4$ at $N_c=3$ and using FDA, we obtain  
\begin{equation}
    \begin{split}
        \frac{dN^{(4)}_{\rm sym}}{d\eta d\v{q}_1^2 d\xi d\v{q}_2^2}
        &
          =\frac{16e^2e_f^2S_{\perp}}{27\pi^4} \alpha^2_s (N_c^2+1)
          (\mathcal{W}_1+\mathcal{W}_2+\mathcal{W}_3)
         \end{split}
\end{equation}
where $\mathcal{W}_i$ are the following integrals (although our equations are valid for  $N_c=3$ only, we keep explicit variable $N_c$ in order to distinguish different contributions) 
\begin{equation}
\label{Eq:W1}
    \begin{split}
      \mathcal{W}_1&=(N_c^2-1) \int_{\v k,\v{\bar{k}}}\Gamma(\v{q}_1,\v{q}_1+\v{k},\v{q}_1+\v{k})\Gamma(\v{q}_2,\v{q}_2+\v{\bar{k}},\v{q}_2+\v{\bar{k}})     D(\v{k})D(\v{\bar{k}})    
        \\
       &\times\Big(2\mathcal{I}_{\varepsilon_f}(| \v{q}_2 +\v{\bar{k}}|)+2\mathcal{I}_{\varepsilon_f}(|\v{q}_1+\v{k}|)
       - \\& \quad -\mathcal{I}(\varepsilon_f,|\v{q}_2-\v{q}_1
       +\v{\bar{k}}-\v{k}|)-\mathcal{I}_{\varepsilon_f}(|\v{q}_1+\v{q}_2+\v{k}+\v{\bar{k}}|)\Big) \,,
    \end{split}
\end{equation}
\begin{equation}
\label{Eq:W2}
    \begin{split}
      \mathcal{W}_2&= \int_{\v k,\v{\bar{k}}}\Gamma(\v{q}_1,\v{q}_1+\v{k},\v{q}_1-\v{\bar{k}})\Gamma(\v{q}_2,\v{q}_2-\v{k},\v{q}_2+\v{\bar{k}})    D(\v{k})D(\v{\bar{k}}) 
        \\
       &
       \times\Big({\mathcal{I}_{\varepsilon_f}(|\v{q}_2-\v{k}|)+\mathcal{I}_{\varepsilon_f}(|\v{q}_1+\v{k}|)+\mathcal{I}_{\varepsilon_f}(|\v{q}_2+\v{\bar{k}}|)+
        \mathcal{I}_{\varepsilon_f}(|\v{q}_1-\v{\bar{k}}|)}\\&-\mathcal{I}_{\varepsilon_f}(|\v{q}_1+\v{q}_2|)
        -\mathcal{I}_{\varepsilon_f}(|\v{k}+\v{\bar{k}}|)
        -\mathcal{I}_{\varepsilon_f}(|\v{q}_2-\v{q}_1 + \v{\bar{k}}-\v{k}|)
        \Big)\,,
    \end{split}
\end{equation}
\begin{equation}
\label{Eq:W3}
    \begin{split}
      \mathcal{W}_3&= \int_{\v k,\v{\bar{k}}}\Gamma(\v{q}_1,\v{q}_1+\v{k},\v{q}_1+\v{\bar{k}})\Gamma(\v{q}_2,\v{q}_2+\v{\bar{k}},\v{q}_2+ k)    D(\v{k})D(\v{\bar{k}})      
        \\
       &
       \times\Big({\mathcal{I}_{\varepsilon_f}(|\v{q}_2 + \v{\bar{k}}|)
    +\mathcal{I}_{\varepsilon_f}(|\v{q}_1 +\v k|)+\mathcal{I}_{\varepsilon_f}(|\v{q}_1 + \v{\bar{k}}|)+
        \mathcal{I}(\varepsilon_f,|\v{q}_2 +\v k|}\\
        &-\mathcal{I}_{\varepsilon_f}(|\v{q}_2-\v{q}_1|)
    -\mathcal{I}_{\varepsilon_f}(|\v{k}-\v{\bar{k}}|)-\mathcal{I}_{\varepsilon_f}(|\v{q}_1+\v{q}_2 + \v{k}+\v{\bar{k}}|)\Big)\,.
    \end{split}
\end{equation}
In these contributions one can identify the underlying physical origin: 
\begin{itemize}
    \item Uncorrelated production:
    the terms appearing in the second line  of Eq.~\eqref{Eq:W1} and proportional to 
    $\mathcal{I}(\varepsilon_f,|\v{q}_1 +\v k|$ and $\mathcal{I}_{\varepsilon_f}(|\v{q}_2 +\v{\bar{k}}|)$ 
    represent  {uncorrelated production}. This is 
    manifested by factorization of the momenta $\v{q}_1$ and $\v{q}_2$ as well  as the integral with respect to  $\v k$ and $\v{\bar{k}}$.

    \item Bose Enhancement in the projectile: 
    the last line of Eq.~\eqref{Eq:W1} is due to Bose Enhancement in the projectile wave function. The two gluons from the projectile scatter on the target independently.
    This contribution is peaked at the momenta when the emitted gluons from the projectile have 
    either colinear or anti-colinear momenta, that is 
    when the arguments of the $\mathcal{I}$ function vanish. There is also an $N_c^2$-suppressed corrections to this contribution in the last terms of  Eqs.~\eqref{Eq:W2} and ~\eqref{Eq:W3}.

    \item Hanbury Brown and Twiss effect: the terms $\mathcal{I}(\varepsilon_f,|\v{q}_1 \pm \v{q}_2| )$ originate from  the HBT effect.  
    
       \item Bose Enhancement in the target: 
    Gluons originating from the target and having  
    either colinear or anti-colinear momenta lead to the Bose Enhancement  in target wave function. This is the origin of the rest of terms in the above expression. This is obvious for the contribution is proportional to $\mathcal{I}_{\varepsilon_f}(|\v k\pm\v{\bar{k}}|)$. The rest of the contributions are also due to Bose enhancement of the target: they correspond to the situation where the two gluons from the projectile scatter in the correlated manner from the target-by exchanging either the same or opposite transverse momenta. The somewhat unusual momentum combinations in factors of ${\cal I}$, like $\mathcal{I}_{\varepsilon_f}(|\v{\bar{k}}-\v{q}_1|)$ express the fact that gluons with momentum of order of inverse dipole size are emitted from the dipole in coherent fashion.

    {%\color{red} 
    Interestingly from Eq.~\eqref{Eq:W1},  one observes that Bose enhancement in the projectile is parametrically of the same order in $N_c$ as the uncorrelated production. This is very different compared to the MV model for the projectile, see Eq.~\eqref{Eq:MVdgp}, where Bose enhancement in the projectile  is suppressed relative to the uncorrelated production by the factor of  $(N_c^2-1)^{-1}$. This difference is due to the fact in the dipole model of the quasi-real photon the $q\bar q$ pair is strongly correlated in color ($q\bar q$ is always in the single state). On the other hand in MV model the valence degrees of freedom are assumed to be completely uncorrelated. This difference in the initial color structure leads to different parametric dependence of the correlated production in the two models.  }
 
  %  \begin{equation}
   %     \begin{split}
    %        &\int_{\v k \v{\bar{k}}}D(-\v{k})D(-\v{\bar{k}}) \times \Big[\Gamma(\v{q}_1,\v{q}_1+\v{k},\v{q}_1-\v{\bar{k}})\Gamma(\v{q}_2,\v{q}_2-\v{k},\v{q}_2+\v{\bar{k}})\Big(\mathcal{I}_{\varepsilon_f}(|-\v{k}+\v{q}_2|)
     %       \\
      %      &
       %     +\mathcal{I}_{\varepsilon_f}(|\v{k}+\v{q}_1|)+\mathcal{I}_{\varepsilon_f}(|\v{\bar{k}}+\v{q}_2|) +
        %\mathcal{I}_{\varepsilon_f}(|-\v{\bar{k}}+\v{q}_1|)\Big)\\
        %&
        %+\Gamma(\v{q}_1,\v{q}_1+\v{k},\v{q}_1+\v{\bar{k}})\Gamma(\v{q}_2,\v{q}_2+\v{\bar{k}},\v{q}_2+ k)\Big(\mathcal{I}_{\varepsilon_f}(|\v{\bar{k}}+\v{q}_2|)
    %+\mathcal{I}_{\varepsilon_f}(|\v{k}+\v{q}_1|)\\&+\mathcal{I}_{\varepsilon_f}(|\v{\bar{k}}+\v{q}_1|)+
     %   \mathcal{I}_{\varepsilon_f}(| \v{k}+\v{q}_2|)\Big)\Big]
      %  \end{split}
    %\end{equation}

\end{itemize}

The remaining, non-symmetric part has kinematic factors that cannot be combined to Lipatov vertices. 
We again write it as a sum of three contributions 
\begin{equation}
    \begin{split}
       \frac{dN_{\rm ns}^{(4)}}{d\eta d\v{q}_1^2 d\xi d\v{q}_2^2}\approx
        & \frac{32 e^2e_f^2}{3 \pi^4}  \alpha_s^2    S_\perp N_c\Big( W_1+ W_2+ W_3\Big)
    \end{split}
\end{equation}
with
    \begin{equation}
        \begin{split}
          W_1
        =&\frac{(N_c^2-1)}{12}\int_{\v k,\v{\bar{k}}}\Big[\Gamma(\v{q}_1,-(\v{q}_1+\v{k}),-(\v{q}_1+\v{k}))+2 \frac{(\v{q}_1+ \v k) \cdot \v{q}_1}{(\v{q}_1+ \v k)^2\v{q}_1^2}  \Big]\Gamma(\v{q}_2,\v{q}_2+\v{\bar{k}},\v{q}_2+\v{\bar{k}})\\
        &\times       D(\v{k})D(\v{\bar{k}})
           \Big(2\mathcal{I}_{\varepsilon_f}(|\v{q}_2+\v{\bar{k}}|)+2\mathcal{I}_{\varepsilon_f}(|\v{q}_1+\v{k}|)
        -\mathcal{I}_{\varepsilon_f}(|\v{q}_1-\v{q}_2+\v{k}-\v{\bar{k}}|)\\&\quad\quad\quad\quad-\mathcal{I}_{\varepsilon_f}(|\v{q}_1+\v{q}_2+\v{k}+\v{\bar{k}}|)\Big)\\
            &
      +
      \frac{(N_c^2-1)}{8}\int_{\v k,\v{\bar{k}}}\Gamma(\v{q}_2,\v{q}_2+\v{\bar{k}},\v{q}_2+\v{\bar{k}})\Gamma(\v{q}_1,-(\v{q}_1+\v{k}),-(\v{q}_1+\v{k}))D(\v{k})D(\v{\bar{k}})
           \\
           &\times \Big(2\mathcal{I}_{\varepsilon_f}(|\v{q}_2+\v{\bar{k}}|)-2\mathcal{I}_{\varepsilon_f}(|\v{q}_1+ \v k|)
        +\mathcal{I}_{\varepsilon_f}(|\v{q}_1-\v{q}_2+\v{k}-\v{\bar{k}}|)+\mathcal{I}_{\varepsilon_f}(|\v{q}_1+\v{q}_2+\v{k}+ \v{\bar{k}}|)\Big)
        \end{split}
    \end{equation}
    \begin{equation}
        \begin{split}
           W_2=&\frac{1}{12}\int_{\v k,\v{\bar{k}}}\Gamma(\v{q}_2,\v{q}_2- k,\v{q}_2+\v{\bar{k}})  \Gamma(\v{q}_1,\v{q}_1+\v{k},\v{q}_1-\v{\bar{k}}) D(\v{k})D(\v{\bar{k}})        \\
        &
        \times \Big[\Big(\mathcal{I}_{\varepsilon_f}(|\v{q}_2-\v{k}|)+\mathcal{I}_{\varepsilon_f}(|\v{q}_1+\v{k}|)+\mathcal{I}_{\varepsilon_f}(|\v{q}_2+\v{\bar{k}}|)+
        \mathcal{I}_{\varepsilon_f}(|\v{q}_1-\v{\bar{k}}|)\\
        &
        -\mathcal{I}_{\varepsilon_f}(|\v{q}_1-\v{q}_2-\v{\bar{k}}+\v{k}|)-\mathcal{I}_{\varepsilon_f}(|\v{q}_1+\v{q}_2|)\Big)\Big] 
        \end{split}
    \end{equation}
    \begin{equation}
        \begin{split}
    W_{3}=&      -  \frac{1}{12} \int_{\v k,\v{\bar{k}}}\Gamma(\v{q}_2,\v{q}_2-\v{\bar{k}},\v{q}_2+\v{k})\Big(\frac{1}{\v{q}_1^2} +\frac{(\v{q}_1+\v{k}) \cdot(\v{q}_1-\v{\bar{k}})}{(\v{q}_1+\v{k})^2(\v{q}_1-\v{\bar{k}})^2} +\Gamma(\v{q}_1,-\v{q}_1-\v{k},-\v{q}_1+\v{\bar{k}}) \Big) 
        \\
        &
        \times D(\v{k})D(\v{\bar{k}})
        \Big(\mathcal{I}_{\varepsilon_f}(|\v{q}_2-\v{\bar{k}}|)+\mathcal{I}_{\varepsilon_f}(|\v{q}_1+\v{k}|)+\mathcal{I}_{\varepsilon_f}(|\v{q}_2+\v{k}|)+
        \mathcal{I}_{\varepsilon_f}(|\v{q}_1-\v{\bar{k}}|)\\&-\mathcal{I}_{\varepsilon_f}(|\v{k}+\v{\bar{k}}|)-\mathcal{I}_{\varepsilon_f}(|\v{q}_1-\v{q}_2|)-\mathcal{I}_{\varepsilon_f}(|\v{q}_1+\v{q}_2+\v{k}-\v{\bar{k}}|)\Big)\\
        &
        -\frac{1}{8}\int_{k,\v{\bar{k}}}\Gamma(\v{q}_2,\v{q}_2-\v{\bar{k}},\v{q}_2+\v{k})\Gamma(\v{q}_1,-\v{q}_1-\v{k},-\v{q}_1+\v{\bar{k}})D(\v{k})D(\v{\bar{k}})\\&
       \times \Big(\mathcal{I}_{\varepsilon_f}(|\v{q}_2-\v{\bar{k}}|)-\mathcal{I}_{\varepsilon_f}(|\v{q}_1+\v{k}|) 
       +\mathcal{I}_{\varepsilon_f}(|\v{q}_2+\v{k}|)
    -\mathcal{I}_{\varepsilon_f}(|\v{q}_1-\v{\bar{k}}|)\\&-\mathcal{I}_{\varepsilon_f}(|\v{k}+ \v{\bar{k}}|)+\mathcal{I}_{\varepsilon_f}(|\v{q}_1-\v{q}_2|)+\mathcal{I}_{\varepsilon_f}(|\v{q}_1+\v{q}_2+\v{k}-\v{\bar{k}}|)\Big)\,. 
        \end{split}
    \end{equation}
We will use this equations for the numerical evaluations.

\section{Inclusive two gluon production: the MV model.}\label{sec5}
In this section  we compute two gluon production by approximating the wave function of the (almost) real photon using the MV model. Note that although we use MV model, we do not discard terms which are suppressed by powers of color charge density, and which are customarily neglected in the CGC literature. 

We start with $\Sigma^{\rm MV}_2$, which is trivial to compute, 
\begin{equation}
\begin{split}
\Sigma^{\rm MV}_2(\v u_1, \v u_2, \bar{\v u}_1,\bar{\v u}_2 )&= 
4\int d^2\bm{x} \mu^2(\bm{x}) f^i(\bar{\v u}_1-\bm{x})f^i(\v u_1-{\bm{x}})f^j(\bar{\v u}_2-\bar{\v u}_1)f^j(\v u_2-\v u_1)
\\
&\left<\Tr \left[ U_{\v u_1}T^{a}[U^{\dagger}_{\v u_2}-U^{\dagger}_{\v u_1}] [U_{\bar{\v u}_2}-U_{\bar{\v u}_1}] T^{a}U^{\dagger}_{\bar{\v u}_1}\right]  \right>_{T}
\end{split}
\end{equation}
This expression modulo the projectile wave function is very much the same as the one we obtained in the previous section, see Eq.~\eqref{Eq:N2}.

The contribution in $\Sigma_3^{\rm MV}$ involves two terms, both of which are of the same order in color charge density as $\Sigma^{\rm MV}_2$,
\begin{equation}
    \begin{split}
       \Sigma^{\rm MV}_3(\v u_1, \v u_2, \bar{\v u}_1,\bar{\v u}_2 )&
       = 2\int d^2\bm{x}\mu^2(\bm{x})f^j(\bar{\v u}_2-\bar{\v u}_1)f^i(\bar{\v u}_1-\bm{x})f^i(\v u_1-\bm{x})f^j(\v u_2-\bm{x})\\
       &
       \Tr\left[  [  U_{\bm{x}}+U_{\v u_1} ] T^a [U^{\dagger}_{\bm{x}}-U^{\dagger}_{\v u_2}] [U_{\bar{\v u}_2}-U_{\bar{\v u}_1}] T^{a} U^{\dagger}_{\bar{\v u}_1}]\right] \\
       &
       +2\int  d^2\bm{x}\mu^2(\bm{x})f^j(\v u_2-\v u_1)f^i(\v u_1-\bm{x})f^i(\bar{\v u}_1-\bm{x})f^j(\bar{\v u}_2-\bm{x})\\
       &
       \Tr \left[ U_{\v u_1} T^{a} [U^{\dagger}_{\v u_2}-U^{\dagger}_{\v u_1}][U_{\bm{x}}-U_{\bar{\v u}_2}]T^a [U^{\dagger}_{\bm{x}}+U^{\dagger}_{\bar{\v u}_1}]\right]
    \end{split}
\end{equation}

Finally, $\Sigma^{\rm MV}_4$ has two contributions, which can be organized by  the power of $\mu^2$. We refer to $\mu^2$ contribution as quantum and $\mu^4$ as classical. We start with the former. Using Eq.~\eqref{Eq:rho4mv}
we obtain 
\begin{equation}
    \begin{split}
        \Sigma_{4,Q}^{\rm MV}(\v u_1, \v u_2, \bar{\v u}_1,\bar{\v u}_2 )
&
=\int d^2\bm{x}\mu^2(\bm{x}) f^i(\bar{\v u}_1-\bm{x})f^j(\bar{u}_2-\bm{x})f^i(\v u_1-\bm{x})f^j(u_2-\bm{x})\\
&
\Big(\Tr( [U^{\dagger}_{\v u_1}-U^{\dagger}_{\bm{x}}][U_{\bar{\v u}_1}-U_{\bm{x}}]T^a[U^{\dagger}_{\bm{x}}-U^{\dagger}_{\bar{\v u}_2}][U_{\bm{x}}-U_{\v u_2}]T^a)\\
&
-\frac13\Tr([U^{\dagger}_{\v u_1}+U^{\dagger}_{\bm{x}}][U_{\bar{\v u}_1}+U_{\bm{x}}]T^a)\Tr( [U^{\dagger}_{\bm{x}}-U^{\dagger}_{\v u_2}][U_{\bm{x}}-U_{\bar{\v u}_2}]T^a)\\
&
-\frac13\tr\Big([U^{\dagger}_{\v u_1}+U^{\dagger}_{\bm{x}}][U_{\bar{\v u}_1}+U_{\bm{x}}]T^a  [U^{\dagger}_{\bm{x}}-U^{\dagger}_{\v u_2}][U_{\bm{x}}-U_{\bar{\v u}_2}]T^a \Big)
    \end{split}
\end{equation}
The classical part  is dominant at large color charge density in the projectile and reads 
\begin{equation}
\label{Eq:54}
    \begin{split}
        \Sigma_{4,C}^{\rm MV}(\v u_1, \v u_2, \bar{\v u}_1,\bar{\v u}_2 )&=4\int d^2\bm{x}d^2\bm{y}\int d^2\bar{\bm{x}}d^2\bar{\bm{y}}f^i(\bar{\v u}_1-\bm{x})f^j(\bar{\v u}_2-\bm{y})f^i(\v u_1-\bar{\bm{x}})f^j(\v u_2-\bar{\bm{y}})\\
& 
\times \left<   \left[[U^{\dagger}_{\v u_1}-U^{\dagger}_{\bar{\bm{x}}}][U_{\bar{\v u}_1}-U_{\bm{x}}]\right]_{e'e} \left[ [U^{\dagger}_{ \v u_2}-U^{\dagger}_{\bar{\bm{y}}}][U_{\bar{\v u}_2}-U_{\bm{y}}]\right]_{c'c}                  \right>_T \\&\times \Big(\delta^{c'e'}\delta^{ce}\mu^2(\bm{x})\mu^2(\bar{\bm{x}})\delta^{(2)}(\bm{x}-\bm{y})\delta^{(2)}(\bar{\bm{x}}-\bar{\bm{y}})\\
&
\quad+\delta^{c'e}\delta^{e'c}\mu^2(\bm{x})\mu^2(\bm{y})\delta^{(2)}(\bm{x}-\bar{\bm{y}})\delta^{(2)}(\bar{\bm{x}}-\bm{y})\\&\quad +\delta^{c'c}\delta^{e'e}\mu^2(\bm{x})\mu^2(\bm{y})\delta^{(2)}(\bm{x}-\bar{\bm{x}})\delta^{(2)}(\bm{y}-\bar{\bm{y}}) \Big)
    \end{split}
\end{equation}
Here the first term in the classification of Ref.~\cite{Kovchegov:2012nd} 
corresponds to the ``square'' diagrams involving the quadrupole, the second to ``cross'' diagrams. The last contribution represents emission where the correlations can only originate from the correlations in the target; this term also gives the leading contribution to the uncorrelated production.

\subsection{Momentum space and FDA}
As in the dipole model, we convert our expressions to momentum space and average over target using  FDA. 
We get 
\begin{equation}
    \begin{split}
       \frac{dN_2}{d\v q^2_1d\v q^2_2 d\eta d\xi}&=
      \frac{4 g^4}{(2\pi)^4}\int_{\v{k}_1,\v{k}_2,\v{k}_3,\v{k}_4}\mu^2(\v{k}_2+\v{k}_4-\v{k}_1-\v{k}_3)\left<\Tr \left[ U_{\v{k}_1}T^{a}U^{\dagger}_{\v{k}_2} U_{\v{k}_3} T^{a}U^{\dagger}_{\v{k}_4}\right]  \right>_{T}\\
       &L^i(\v{q}_2,\v{q}_2+\v{k}_2)L^j(\v{q}_2,\v{q}_2+\v{k}_3) \frac{(\v{q}_1+\v{q}_2+\v{k}_3-\v{k}_4)^i(\v{q}_1+\v{q}_2+\v{k}_2-\v{k}_1)^j}{(\v{q}_1+\v{q}_2+\v{k}_3-\v{k}_4)^2(\v{q}_1+\v{q}_2+\v{k}_2-\v{k}_1)^2}
    \end{split}
\end{equation}
Using FDA, this simplifies into
\begin{equation}
    \begin{split}
         \frac{dN_2}{d\v q^2_1d\v q^2_2 d\eta d\xi}&=\frac{4N_cg^4}{(2\pi)^4}\int d^2 \v x\mu^2(\v x) \int_{\v{k}_1,\v{k}_2} D(\v{k}_1+\v{q}_1)D(-\v{k}_2+\v{q}_2)\\
         &\Big\{ (N_c^2-1) \left(\frac{(\v{k}_2-\v{k}_1) \cdot  L(\v{q}_2,\v{k}_2)}{(\v{k}_2-\v{k}_1)^2}\right)^2
        \\ & \quad  -\frac{(\v{k}_2-\v{k}_1)\cdot L(\v{q}_2,\v{k}_2)}{(\v{k}_2-\v{k}_1)^2}\frac{(\v{k}_2-\v{k}_1) \cdot L(\v{q}_2,\v{q}_2-\v{k}_2-\v{k}_1)}{(\v{k}_2-\v{k}_1)^2}\Big\}
    \end{split}
\end{equation}
%\begin{equation}
 %   \begin{split}
  %     \Sigma^{\rm MV}_3(\v u_1, \v u_2, \bar{\v u}_1,\bar{\v u}_2 )&
   %    = 2\int d^2\bm{x}\mu^2(\bm{x})f^j(\bar{\v u}_2-\bar{\v u}_1)f^i(\bar{\v u}_1-\bm{x})f^i(\v u_1-\bm{x})f^j(\v u_2-\bm{x})\\
    %   &
     %  \Tr\left[  [  U_{\bm{x}}+U_{ \v u_1} ] T^a [U^{\dagger}_{\bm{x}}-U^{\dagger}_{\v u_2}] [U_{\bar{\v u}_2}-U_{\bar{ \v u}_1}] T^{a} U^{\dagger}_{\bar{\v u}_1}]\right] \\
      % &
      % +2\int  d^2\bm{x}\mu^2(\bm{x})f^j( \v u_2-\v u_1)f^i(\v u_1-\bm{x})f^i(\bar{\v u}_1-\bm{x})f^j(\bar{\v u}_2-\bm{x})\\
       %&
      % \Tr \left[ U_{\v u_1} T^{a} [U^{\dagger}_{\v u_2}-U^{\dagger}_{\v u_1}][U_{\bm{x}}-U_{\bar{\v u}_2}]T^a [U^{\dagger}_{\bm{x}}+U^{\dagger}_{\bar{\v u}_1}]\right]
%    \end{split}
%\end{equation}
Similarly for  $\Sigma_3$ contribution we obtain 
\begin{equation}
    \begin{split}
         \frac{dN_3}{d\v q^2_1d\v q^2_2 d\eta d\xi}=& \frac{4 g^4}{(2\pi)^4}\int_{\v{k}_1,\v{k}_2,\v{k}_3,\v{k}_4} \mu^2\Big(\v{k}_4+\v{k}_2-\v{k}_1-\v{k}_3\Big) \Tr[U_{\v{k}_1}T^aU^{\dagger}_{\v{k}_2}U_{\v{k}_3}T^aU^{\dagger}_{\v{k}_4}]\\
    &\Gamma(\v{q}_2,\v{q}_2+\v{k}_2,\v{q}_2+\v{k}_3)L^i(\v{q}_1,\v{k}_1-\v{q}_1)\frac{(\v{k}_4-\v{k}_3-\v{q}_1-\v{q}_2)^i}{(\v{k}_4-\v{k}_3-\v{q}_1-\v{q}_2)^2}
    \end{split}
\end{equation}
where we assumed that $\mu^2(-\v{k}) = \mu^2(\v{k})$. 
The FDA leads to  
\begin{equation}
    \begin{split}
        \frac{dN_3}{d\v q^2_1d\v q^2_2 d\eta d\xi} 
        =&\frac{4N_cg^4}{(2\pi)^4}\int d^2 \v x\mu^2(\v x) \int_{\v{k}_1,\v{k}_2} D(\v{k}_1+\v{q}_1)D(-\v{k}_2+\v{q}_2)\\
        &\Big\{(N_c^2-1) L^i(\v{q}_1,\v{k}_1)\frac{(\v{k}_1-\v{k}_2)^i}{(\v{k}_1-\v{k}_2)^2}\Gamma(\v{q}_2,\v{k}_2,\v{k}_2)\\& \quad -L^i(\v{q}_1,\v{k}_1)\frac{(\v{k}_1-\v{k}_2)^i}{(\v{k}_1-\v{k}_2)^2}\Gamma(\v{q}_2,\v{k}_2,\v{q}_2-\v{q}_1-\v{k}_1)\Big\}\,.
    \end{split}
\end{equation}

Finally the $\Sigma_4$ contribution is given by two terms proportional to  $\mu^2$ and $\mu^4$. Starting with the former, in FDA we obtain 
\begin{equation}
    \begin{split}
         &\frac{dN_4^Q}{d\v q^2_1d \v q^2_2 d\eta d\xi}\\
         =&\frac{g^4N_c}{(2\pi)^4}\int_{\v k,\v{\bar{k}}} D(-\v{k})D(-\v{\bar{k}})\int d^2\bm{x}\mu^2(\bm{x})\\&
         \Big\{(N_c^2-1)\Big[\Gamma(\v{q}_1,\v{q}_1+\v{k},\v{q}_1+\v{k})-\frac13\Gamma(\v{q}_1,-\v{q}_1-\v{k},-\v{q}_1-\v{k}) \Big]\Gamma(\v{q}_2, \v{q}_2+ \v{\bar{k}},\v{q}_2+\v{\bar{k}})\\
         &+\frac23\Gamma(\v{q}_1,-\v{q}_1-\v{k},-\v{q}_1-\v{\bar{k}})\Gamma(\v{q}_2, \v{q}_2+ \v{\bar{k}},\v{q}_2+\v k)\\
         &+\Big[ \Gamma(\v{q}_1,\v{q}_1+\v{k},\v{q}_1+\v{\bar{k}})  -\frac13\Gamma(\v{q}_1,-\v{q}_1+\v{k},-\v{q}_1-\v{\bar{k}})\Big]   \Gamma(\v{q}_2, \v{q}_2+ \v k,\v{q}_2+\v{\bar{k}})\Big\}
    \end{split}
\end{equation}
The latter is given by 
\begin{equation}
\label{Eq:MVdgp}
    \begin{split}
     \frac{dN_4^C}{d\v q^2_1d \v q^2_2 d\eta d\xi}
     &=\frac{4g^4}{(2\pi)^4} \int_{\v k,\v{\bar{k}}}D(-\v{k})D(\v{\bar{k}})\Big\{\Gamma(\v{q}_1,\v{q}_1+\v{k},\v{q}_1+\v{k})
\Gamma(\v{q}_2,\v{q}_2-\v{\bar{k}},\v{q}_2-\v{\bar{k}})\\&
\left[|\mu^2(0)|^2(N_c^2-1)^2+(N_c^2-1)|\mu^2(\v{q}_1-\v{q}_2+\v{k}+\v{\bar{k}})|^2 \right]\\&
+(N_c^2-1)|\mu^2(\v{q}_1+\v{q}_2+\v{k}+\v{\bar{k}})|^2\Gamma(\v{q}_1,\v{q}_1+\v{k},\v{q}_1+\v{k})\Gamma(\v{q}_2,\v{q}_2+ \v{\bar{k}},\v{q}_2+\v{\bar{k}})\\
&+\Gamma(\v{q}_1,\v{q}_1+\v{k},\v{q}_1+\v{\bar{k}})\Gamma(\v{q}_2,\v{q}_2+ k,\v{q}_2+\v{\bar{k}})\Big( |\mu^2(\v{q}_1+\v{q}_2+\v{k}+\v{\bar{k}})|^2\\&
+(N_c^2-1)|\mu^2(\v{q}_1-\v{q}_2)|^2+|\mu^2(\v{k}-\v{\bar{k}})|^2\Big)\\
&+\Gamma(\v{q}_1,\v{q}_1+\v{k},\v{q}_1+\v{\bar{k}})\Gamma(\v{q}_2,\v{q}_2-\v{k},\v{q}_2-\v{\bar{k}})\Big((N_c^2-1)|\mu^2(\v{q}_1+\v{q}_2)|^2\\&
+|\mu^2(\v{q}_1-\v{q}_2+\v{k}+\v{\bar{k}})|^2+|\mu^2(\v{k}-\v{\bar{k}})|^2\Big)
    \end{split}
\end{equation}

The physical origin of different terms can be easily identified:
\begin{itemize}
    \item Uncorrelated production: the contribution proportional to $(N_c^2-1)^2|\mu^2(0)|^2$ describes uncorrelated two gluon production
    \item Bose Enhancement in the projectile: this term is proportional to $(N_c^2-1)|\mu^2(\v{q}_1 \pm \v{q}_2 +\v{k}+\v{\bar{k}})|^2$. It is suppressed by the factor of $1/N_c^2$ compared to the uncorrelated production as the initial gluons have to have the same color. As in the dipole case, there is a correction to Bose Enhancement in the projectile which is suppressed by a further factor of $1/N_c^2$. It is proportional to   $|\mu^2(\v{q}_1\pm \v{q}_2+\v{k}+\v{\bar{k}} )|^2$. 
    
    \item Bose Enhancement in the target: the terms proportional to  $|\mu^2(\v{k}-\v{\bar{k}} )|^2$ represent Bose Enhancement in the target; they are suppressed by $1/N_c^4$ in the dense regime.

    \item HBT: finally the terms $(N_c^2-1)|\mu^2(\v{q}_1\pm \v{q}_2)|^2$ represent HBT correlation. This term is suppressed by $1/N_c^2$ relative to the uncorrelated production as the color of the final state gluons should be the same. 
    
\end{itemize}

\section{Numerical results}
\label{Sec:Num}
In this section we evaluate numerically inclusive two gluon production in both models for the projectile wave function.

We do not have ambitions to fit experimental data, and we therefore do not try to optimize parameters of the model. Nevertheless we want to be able to have at least a qualitative idea on the ballpark of the CGC predictions for relevant observables.  

The first question we are faced with is that of subtraction of the background.
In experiment, low multiplicity events are used to subtract trivial contribution from high multiplicity data, which is then analyzed in terms of flow harmonics. In our calculation using the MV model for the projectile, this procedure naturally corresponds to subtracting the $\mu^2$ terms as they play a dominant role at low multiplicity and are negligible compared to $\mu^4$ terms at high multiplicity. We thus take as our ``signal''  only the $\mu^4$ contribution, which corresponds to the classical gluon production as defined in the previous section. In the dipole model for the projectile wave function, the distinction between low and high multiplicities is trickier to make since we do not have an explicit parameter $\mu^2$. Nevertheless motivated by the MV model, we will use the symmetric part of the matrix element $\Sigma_4$ as our high multiplicity ``signal''. For $N_c=3$, as we demonstrated in Sect.~\ref{Sect:2glDPM}, it has a compact form and is clearly not dominated by back-to-back production.

For numerical calculations we will use the following set of parameters: the saturation momentum of the target is chosen as $Q_s=2$ GeV, the infrared cut-off entering into the Poisson equation of the MV model is set to be equal to $\Lambda = 200$ MeV, additionally to insure that the logarithm of the argument of the logarithm entering into the adjoint dipole of Eq.~\eqref{Eq:KovnersHammer} does not change sign we use 
\begin{align}
    D(\v r) = \exp\left[-\frac{1}{4}Q_s^2r^2\ln(\frac{1}{\Lambda^2r^2}+e) \right]\,. 
\end{align}
For the nearly real photon wave function, we use a small but non-zero value of $Q$  in order to regularize large distance physics. In our calculations we fix $\epsilon_f = 100$ MeV. This roughly corresponds to $Q\sim 200$ MeV. 

To ensure that the projectile photon has a finite size, 
in the projectile MV model we introduce the projectile size $R$ through the impact parameter dependent parameter $\mu^2(\v x)$:
\begin{align}
    \mu^2(\v x) = {\cal N} \exp{-\frac{\v x^2}{R^2}}\,.
\end{align}
Since our signal is a homogeneous function of $\mu^2$, the  constant $ {\cal N}$ does not play any role in the normalized quantities considered below and we do not need to specify it.   
Here we do not explore the dependence on the shape of the projectile as it has little effect on the observable in this approach, except for unrealistically large ellipticities, see Ref.~\cite{Kovner:2018fxj}.

The code is publicly available in Ref.~\cite{UPCcode}.   

%We illustrate the shape of different contribution to double inclusive production by showing the angular correlation function for each contribution separately, that is 
%\begin{align}
%    C(\Delta \theta) = \frac{  \int d \theta_1 d \theta_2 \delta(\Delta \theta - \theta_2 +\theta_1)   \frac{d N^{(i)}}{d\v q^2_1 d \v q^2_2 d\eta d\xi} } { \int d \theta_1 d \theta_2  \frac{d N^{(i)}}{d\v q^2_1 d \v q^2_2 d\eta d\xi}  }
%\end{align}

In order to compute the two particle $v_2$ as a function of the transverse momentum,  we define the $n$-th Fourier mode of the particle production cross section $V_n(p_\perp)$  
\begin{align}
    V_n(q_1) = 
    \int d \theta_1
    \int_{0}^{p^{\rm max}_\perp} d^2 \v{q}_2 
    \exp(i n  \Delta \theta ) 
    \frac{d N}{d\v q^2_1 d \v q^2_2 d\eta d\xi}
\end{align}
where $\Delta \theta = \theta_2-\theta_1$ is the angle between $\v q_1$ and $\v q_2$. 
Additionally, the integrated Fourier modes are defined according to  
\begin{align}
    V_n = 
    \int_0^{p^{\rm max}_\perp} d^2 \v{q}_1 
    \int_0^{p^{\rm max}_\perp} d^2 \v{q}_2 
    \exp(i n  \Delta \theta ) 
    \frac{d N}{d\v q^2_1 d \v q^2_2 d\eta d\xi}\,.
\end{align}
Then we have 
\begin{align}
\label{Eq:v22}
    v^{(2)}_2(p_\perp) = \sqrt{\frac{V_2(p_\perp)}{V_0(p_\perp)}}
\end{align}
for the second Fourier harmonic of the two particle correlation function~\footnote{%\color{red}
To relate to  notations of Ref.~\cite{ATLAS:2021jhn}, our $(v^{(2)}_2(p_\perp))^2$ is the same as  $v_{2,2}(p_\perp=p_a, p_b)$ of Ref.~\cite{ATLAS:2021jhn} with $p_b$ integrated over the specified momentum range.}. 

Experimental collaborations, in addition to $v^{(2)}_2(p_\perp)$ also extract $v_{2}$ whose definition is based on the assumption of the factorization of the hydrodynamic flow. 
In hydrodynamics, a two-particle azimuthal harmonic $v^{(2)}_2$ is a product of single particle  azimuthal anisotropies, $v_2$. This motivates considering the observable 
\begin{align}
\label{Eq:v2}
    v_{2}(p_\perp) = \frac{V_2(p_\perp)/V_0(p_\perp)}{\sqrt{V_2/V_0}}\,.
\end{align}
Experimentally the two observables are quite close to each other \cite{ATLAS:2021jhn}.
In our calculations, there is no reason at all to expect such hydrodynamics-like factorization. It is nevertheless useful to consider $v_{2}$ in order to get an idea of how much the two quantities differ from each other in our framework. 

\begin{figure}
    \centering
\includegraphics[width=0.49\linewidth]{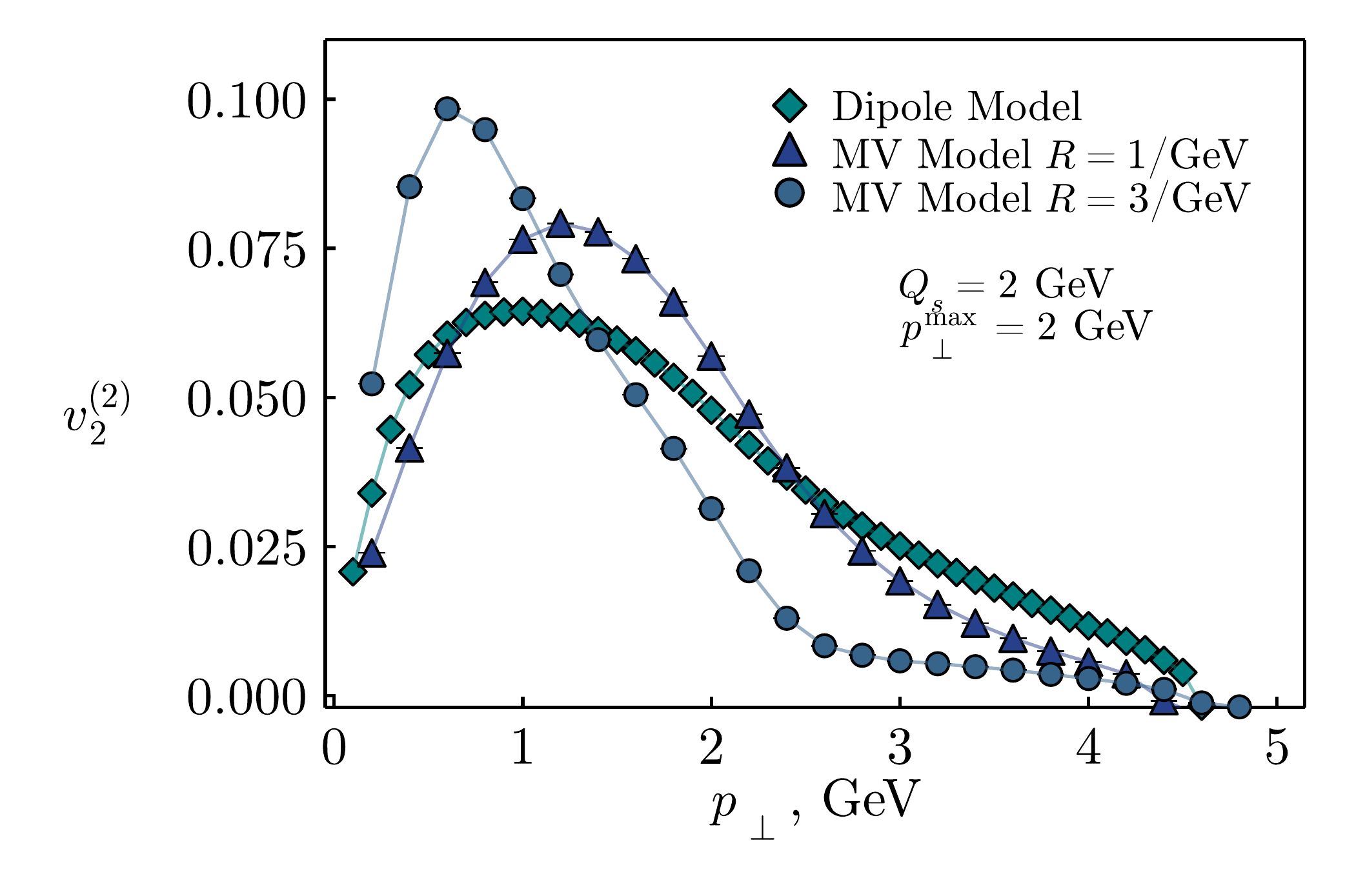}
\includegraphics[width=0.49\linewidth]{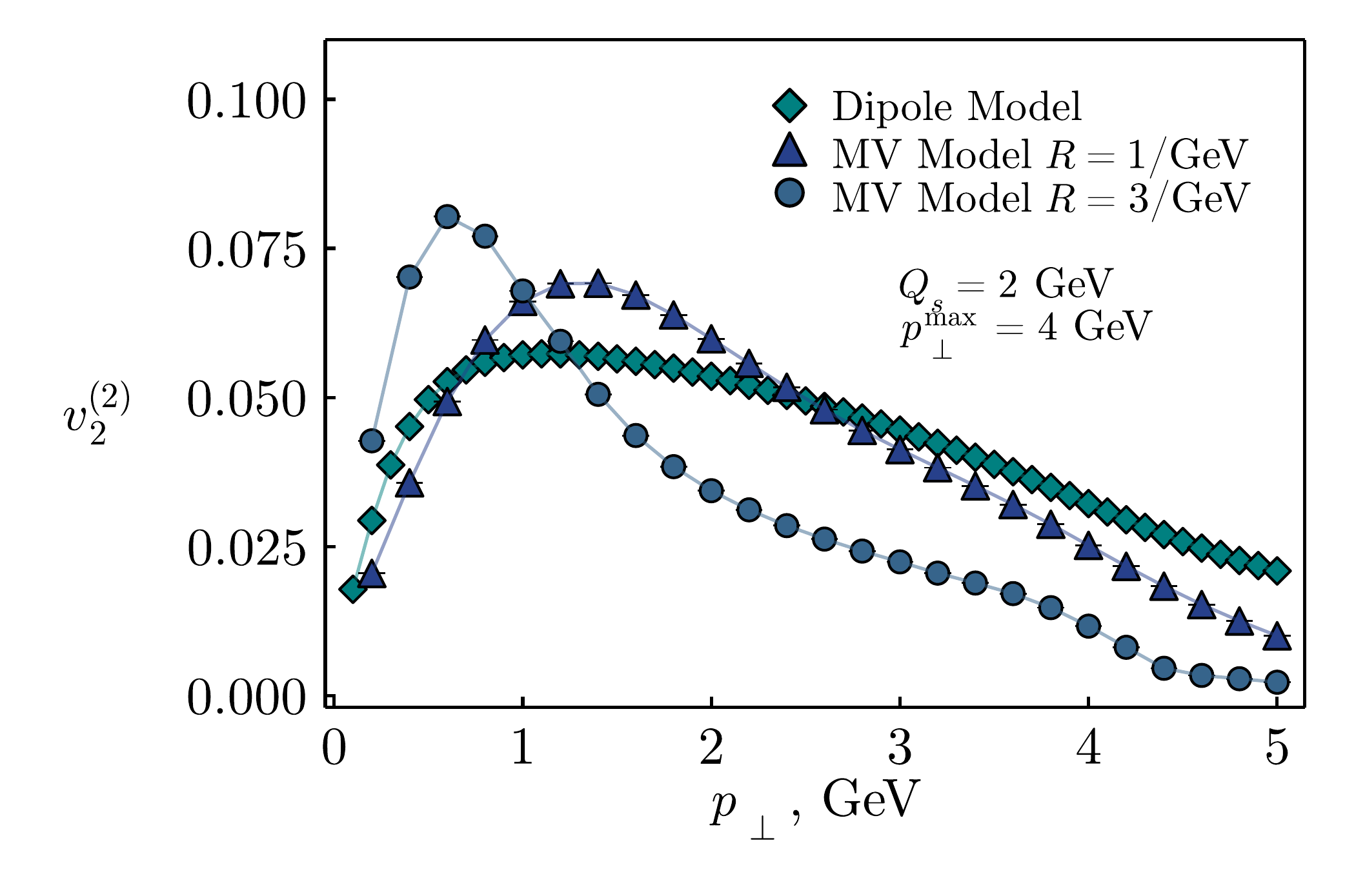}
    \caption{The second Fourier harmonic of the two particle correlation function, $v^{(2)}_2$, see Eq.~\eqref{Eq:v22}, as a function of the transverse momentum of the "trigger" gluon. The momentum of the "associated" gluon is integrated from $0$ to $p_\perp^{\rm max}$.   }
    \label{Fig:v22}
\end{figure}

\begin{figure}
    \centering
\includegraphics[width=0.49\linewidth]{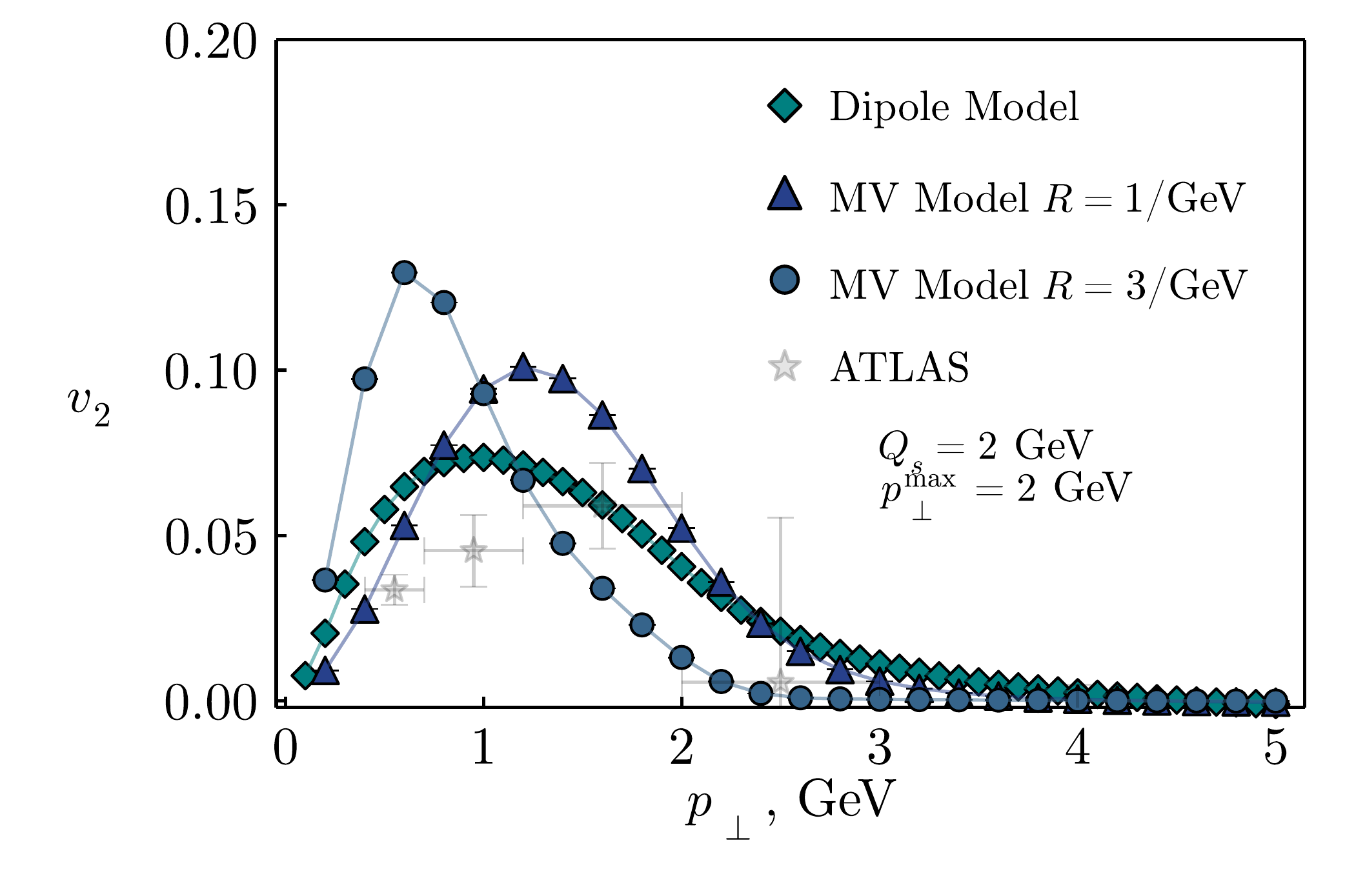}
\includegraphics[width=0.49\linewidth]{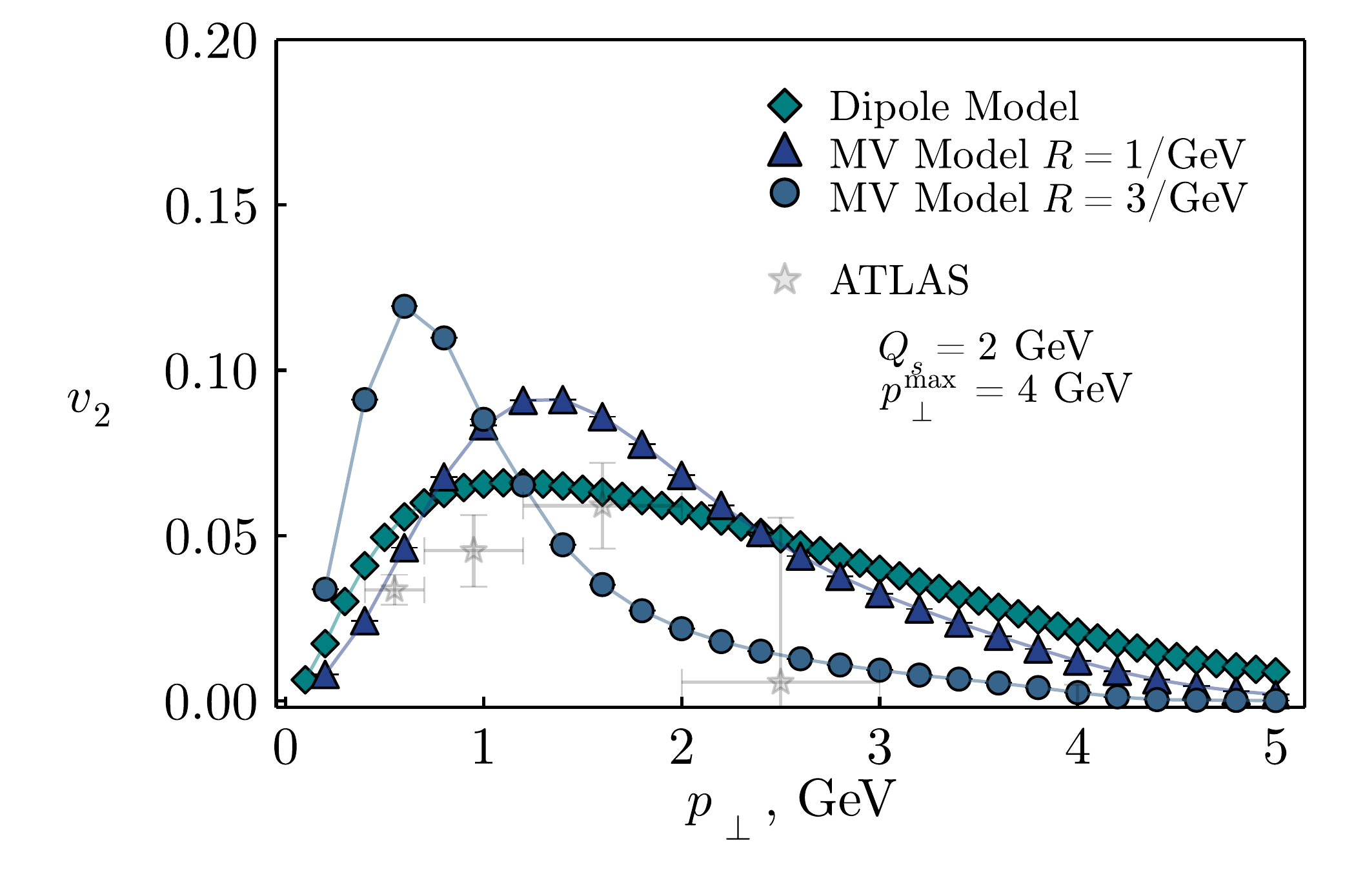}
    \caption{The elliptic flow $v_2$ extracted from two particle correlations function using the definition motivated by factorization, $v_2$, see Eq.~\eqref{Eq:v2}, as a function of the transverse momentum of the "trigger" gluon. The momentum of the associated gluon is integrated from $0$ to $p_\perp^{\rm max}$. To guide the eye we plotted ATLAS data from Ref.~\cite{ATLAS:2021jhn}. }
    \label{Fig:v2}
\end{figure}

We plot these two quantities in Figs.~\ref{Fig:v22} and~\ref{Fig:v2}. To guide the eye we also added experimental data to compare it to $v_{2}$. Note that we did not plot the last experimental point which is located at rather large negative values of $v_2$, see ATLAS~\cite{ATLAS:2021jhn}. 
The calculations are done for two upper limits of the integration with respect to  the transverse momentum $p^{\rm max}_\perp = 2$ and 4 GeV. The former choice is the same as in the ATLAS measurements, while the later choice is motivated by comparison with the study in Ref.~\cite{Shi:2020djm}.

%\begin{center}
 %   \includegraphics[scale=0.6]{Numerics/dipole_everything.pdf}\\[12pt]
%\begin{figure}
 %   \centering
 %\includegraphics[scale=0.6]{Numerics/Dipole.pdf}
  %  \caption{The angular correlations from each separate contribution in the dipole approximation.}
   % \label{Fig:Sigma_dipole}
%\end{figure}

%     \includegraphics[scale=0.6]{Numerics/MV.pdf}\\[12pt]

%\begin{figure}
%    \centering
 %    \includegraphics[scale=0.6]{Numerics/sym.pdf}
 %   \caption{The angular correlation in the symmetric part of the double inclusive gluon production in the dipole approximation and the MV model. The momenta of both gluons is the same, $|\v{q}_1|=|\v{q}_2|=3 $ GeV.}
 %   \label{Fig:Symm_dipole}
%\end{figure}

%\end{center}

\section{Discussion: {%\color{red}:  
factorization and forward production.}}
\label{sec7}
We now discuss our results. 
We stress again that our goal was not to fit the ATLAS data as we do not believe that our calculational framework is constrained enough for such an endeavor. First, we do not include hadronization in our calculation, and that may have a large effect on the correlated part of particle production. In addition the wave function of the real photon has to be modelled in some way, and this modelling leaves a lot of freedom.  Instead we studied two ubiquitous models used in a variety of CGC calculations to understand the qualitative features of the effect.
Consequently we did not try to optimize the parameters of the MV model and the dipole model of the photon when calculating $v_2$. 

Before we proceed, we want to put our study in the perspective by commenting on the existing literature and comment on the comparison of our results to those of Ref.~\cite{Shi:2020djm}. Direct comparison between the two is hard, since the calculational frameworks are very different, even though both calculations are based on the CGC approach. Our calculation is appropriate at non forward rapidities where particle production is dominated by gluons, whereas the authors of Ref.~\cite{Shi:2020djm} use the hybrid approach which restricts their calculation to the photon going direction. One could attempt to take the model used in Ref.~\cite{Shi:2020djm} for valence parton distribution in the photon and use it as yet another model for the photon. This is however not a well posed problem since no color correlations between valence partons have been taken into account in Ref.~\cite{Shi:2020djm} due to collinear approximation invoked there, while such color correlations are clearly important away from forward rapidity especially at relatively low transverse momenta.

There is however one qualitative observation that we can make with confidence. The $v_2$ is our calculation exhibits in all models the same qualitative trend: it rises at low transverse momentum to a maximum at $p_\perp$ around $ 1$ GeV, and drops thereafter, see Figs.~\ref{Fig:v22} and \ref{Fig:v2}. This is also a clear trend in the experimental data. On the other hand $v_2$ calculated in Ref.~\cite{Shi:2020djm} never reaches a maximum, but instead keeps growing at high momentum far beyond the point where experimentally it is observed to fall. Qualitative this difference in behavior is easy to understand. As we will explain below the origin of the turnover in our calculation  is the dominance of a very narrow gluon HBT peak. {%\color{red} 
We stress that the dominance of the HBT peak in correlations at mid rapidity is not an artifact of our approximation. It was demonstrated both, within the FDA in Ref.~\cite{Altinoluk:2018ogz} and also using the lattice simulations of the MV model without invoking  FDA in Ref.~\cite{Kovner:2018fxj}  . 

In contrast to the above, in the calculation of forward production in Ref.~\cite{Shi:2020djm} the gluon HBT effect is simply absent. This calculation treats forward moving partons as distinguishable, and therefore does not account for quantum statistics effect, like Bose enhancement and HBT. Instead the physical origin of the correlations in Ref.~\cite{Shi:2020djm} is the so called color domain structure in the target~\cite{Kovner:2010xk}\footnote{%\color{red}
Note that although color domains are not introduced explicitly in Ref.~\cite{Shi:2020djm}, they are an inherent feature in the  configuration by configuration realization of  the MV model used in the calculations of Ref.~\cite{Shi:2020djm}. }  The color domain effect is the more efficient, the higher the transverse momentum of produced particles. High $p_\perp$ particles are produced nearby in the coordinate space, and therefore are more probable to probe the same domain in the target. Thus the approach of Ref.~\cite{Shi:2020djm} does not contain a mechanism that could lead to decreasing $v_2$ at high momentum\footnote{The color domain structure in principle also contributes to correlations in our approach where it appears in the guise of target BE correlations. However as explained above it is strongly suppressed relative to the gluon HBT. {For example in the MV model at large $N_c$ the target BE correlations are suppressed by $1/N_c^2$ relative to the leading correlated terms. Even if one such sub-leading contribution grows at high $p_\perp$ this does not necessarily result in the growth of the total correlated signal, as there are other contributions of the same order (e.g. sub-leading BE of the projectile corrections). Those are not necessarily positive and may naturally temper this growth. In the kinematic region we consider here the gluon HBT effect is by far the dominant one, and  its behavior  determines the overall behavior of the correlated signal.}  }.}

\begin{figure}
    \centering
\includegraphics[width=0.49\linewidth]{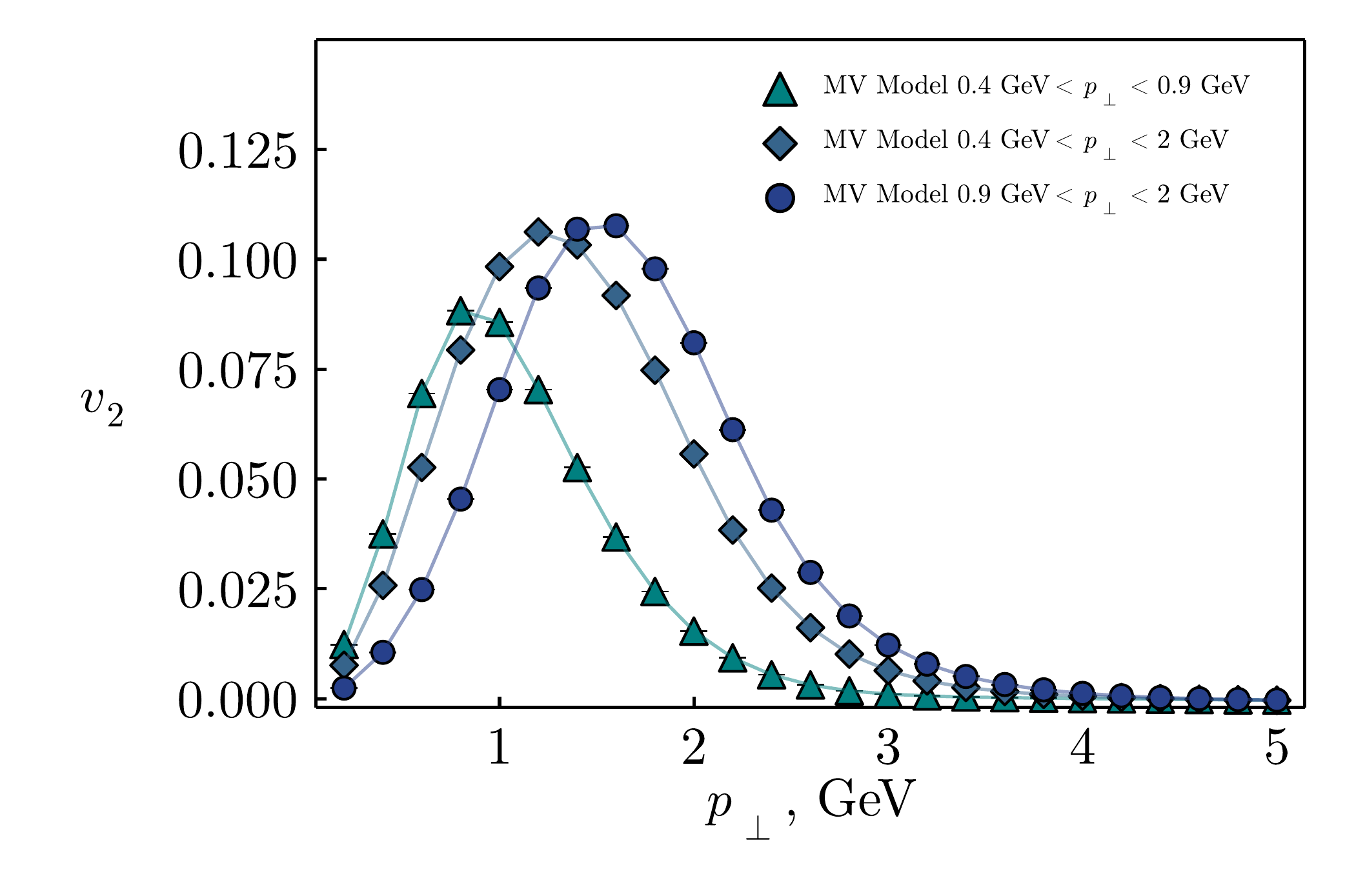}
\includegraphics[width=0.49\linewidth]{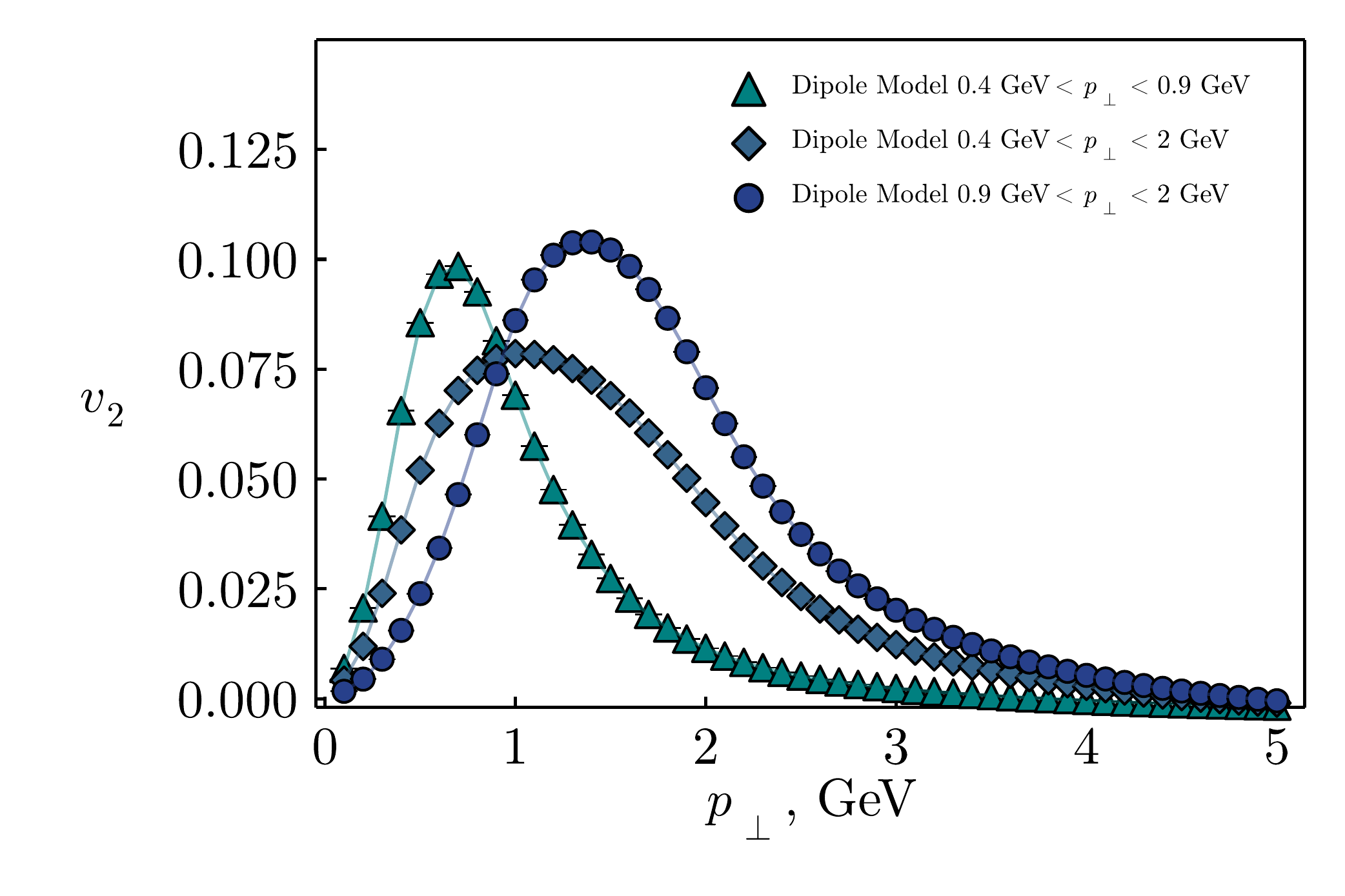}
    \caption{The elliptic flow $v_2$ for three different kinematic ranges of the trigger particle.  Here as in the previous figure,  $Q_s=2$ GeV.  The size of the projectile is set  by $R=1$/GeV.}
    \label{Fig:v21}
\end{figure}

\begin{figure}
    \centering
\includegraphics[width=0.49\linewidth]{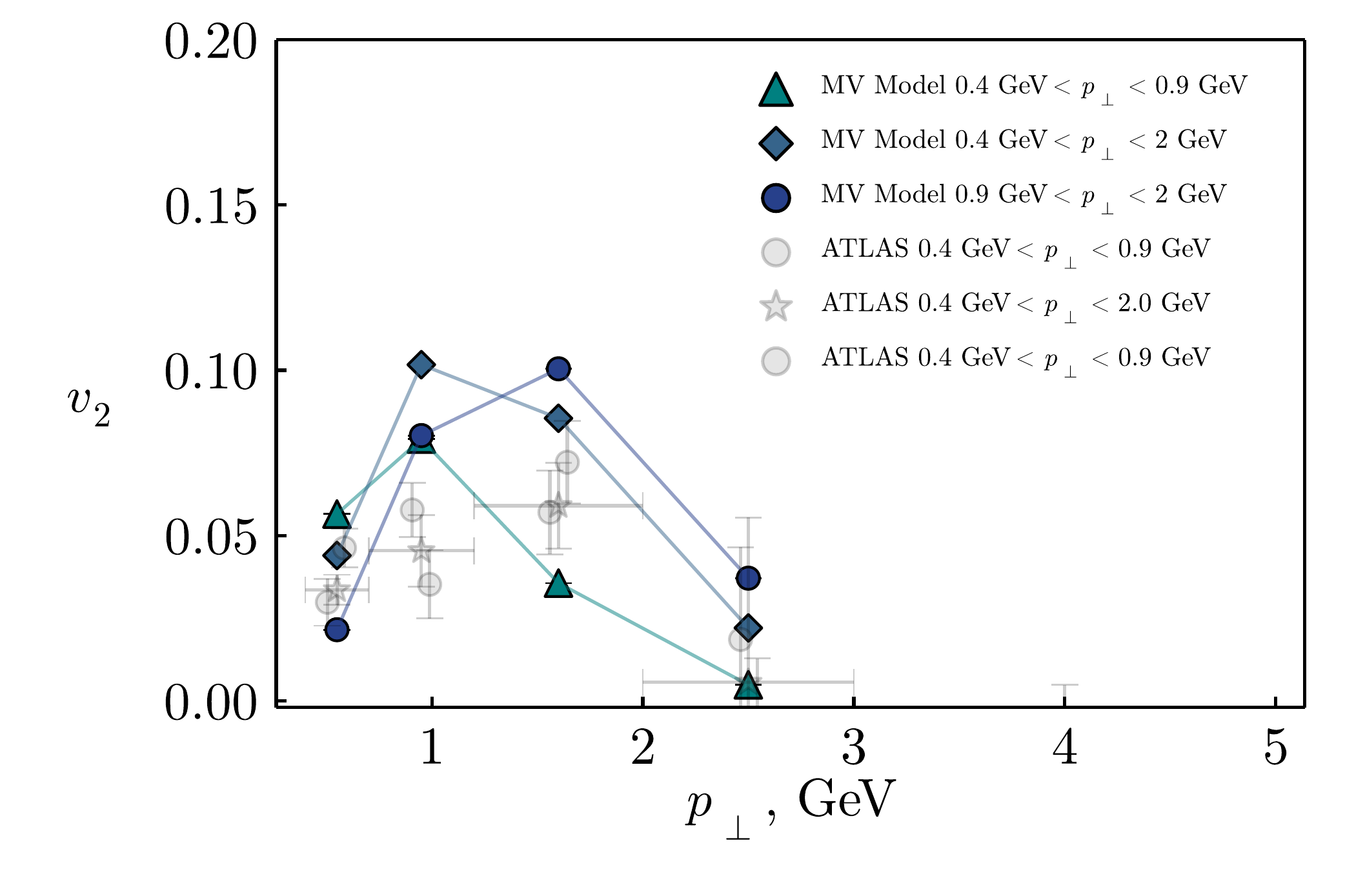}
\includegraphics[width=0.49\linewidth]{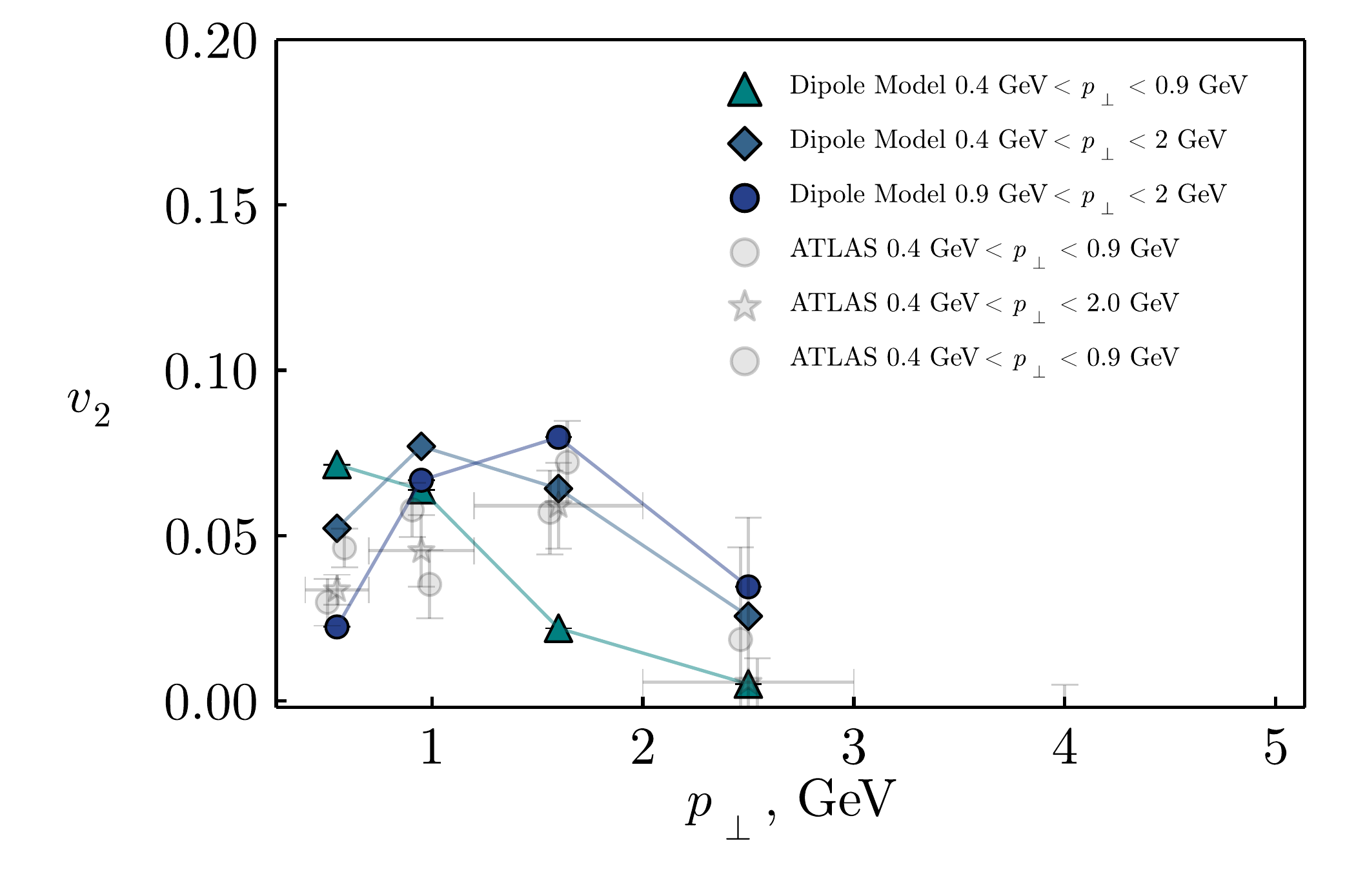}
    \caption{The same as Fig.~\ref{Fig:v21} but binned with the same bin size as the ATLAS analysis. }
    \label{Fig:v23}
\end{figure}

Consider now our numerical results on Figs.~\ref{Fig:v22} and ~\ref{Fig:v2}. The first observation is that both the MV model and the dipole model produce $v_2$ of the same order as the experimental data. The shape of the momentum dependence in the two models is slightly different, with MV mode $v_2$ rather sharply peaked at low momentum.

One somewhat surprising feature of our results is that $v_2$ and $v_2^{(2)}$ turn out to be not very different. Naive expectation is that $v_2$ should drop very quickly once the momentum of the trigger particle is outside the integration interval of the momentum of the associate. The reasoning  for this  is that the two particle correlations in the CGC framework receive a very large contribution from the gluon HBT effect. The HBT peak is very narrow, in fact for a large projectile its width in momentum space is inversely proportional to the area. As a result for a large area projectile  the correlation drops by several orders of magnitude once the momentum of the trigger is outside of the momentum bin of the associate~\cite{Altinoluk:2018ogz}. It is thus natural to expect $V_2/V_0\gg V_2(p_\perp)/V_0(p_\perp)$ and therefore $v_2\ll v_2^{(2)}(p_\perp)$ when $p_\perp>p_\perp^{\rm max}$.
Instead we observe that although $v_2$ does indeed drop somewhat faster for $p_\perp>p_\perp^{\rm max}$, the difference is not that spectacular especially for the dipole model photon. We believe this is due to the finite size of the projectile wave function, which leads to widening of the HBT peak and tempers the drop in the HBT correlation.

With these considerations in mind we  plot the dependence of $v_2$ on the choice of the momentum bin of the associated particle. We choose the bins the same as in the ATLAS data~\cite{ATLAS:2021jhn}. As mentioned above, the hydrodynamic-like universality/{factorization} would suggest that this quantity does not depend on the bin size and location. Our results on the other hand are not expected to display such independence. In Fig.~\ref{Fig:v21} we plot $v_2$ as a function of the momentum of the trigger particle for three different bin choices of the associate. Indeed we observe a rather strong dependence on the bin choice consistent with the discussion above. Namely, $v_2$ is large when the momentum of the trigger particle $p_\perp$ is within the bin, and drops fairly quickly once $p_\perp$ is outside the bin. This behavior is common for the MV model and the dipole model of the photon. At some values of $p_\perp$ the difference between the values of $v_2$ is very large, with ratio between $v_2$ for different bins reaching about 7 or 8 at the maximum. This seems to be in stark contradiction with experimental data. However, one has to keep in mind that the experimental test of universality/{factorization} in Ref.~\cite{ATLAS:2021jhn} was not performed for a fixed momentum of a trigger particle, rather $p_\perp$ was also binned. Following the same procedure we plot on Fig.~\ref{Fig:v23} values of $v_2$ for four bins of $p_\perp$ chosen in the same way as in~Ref.\cite{ATLAS:2021jhn}. Interestingly, the binning has a very strong effect: since the curves in Fig.~\ref{Fig:v21} are all bell shaped on the scale comparable with the bin size, integrating over the $p_\perp$ bins leads to significant reduction of differences in the values of $v_2$. The ratio between the values of $v_2$ for different associate bins is now at most 3 or 4 rather than 7 or 8. This is still significantly larger than for the ATLAS data, where the maximal ratio is closer to 2, but the difference is now not as drastic. Interestingly, the systematics of momentum dependence of our results is quite similar to that in the data, with $v_2$ initially rising, and subsequently falling with $p_\perp$, and $v_2$ for some of the bins crossing over in the vicinity of $1$ GeV. 

We stress again that our purpose in this exercise is not to force-fit the data. It is quite clear that fiddling with the model parameters would allow us to get much closer to experimental points, but it is unclear what it would teach us. Instead, our take home message is twofold. First, we conclude unambiguously that a CGC calculation does not yield results consistent with the hydrodynamic inspired universality of $v_2$, although the values of $v_2$ we obtain are roughly of the same magnitude as the experimental ones. Second, it seems to us that the test of the said universality conducted in Ref.~\cite{ATLAS:2021jhn} is inconclusive. The variation between the different bin values found in Ref.~\cite{ATLAS:2021jhn} albeit broadly consistent with universality, is also quite similar to our (not optimized) CGC results due to binning of the momentum of the trigger particle. A more detailed study of this point seems desirable.

\acknowledgments 
We thank T.~Altinoluk, N.~Armesto, M.~Lublinsky, and A.~Milov for illuminating discussions. 
A.K. is supported by the NSF Nuclear Theory grant 1913890. This material is based
upon work supported by the U.S. Department of Energy, Office of Science, Office of Nuclear
Physics through the Contract No. DE-SC0020081 (H.D. and V.S.).

\bibliography{UPC}
\end{document}